\documentclass[11pt,a4paper,onecolumn,twoside]{rho}
\usepackage[english]{babel}

\doctype{Research Article}
\title{Analytical Nuclear Gradients and Hessians on Quantum Hardware via Orbital-Optimized VQE with Error Mitigation}

\author[{\textasteriskcentered},a]{Renato Olarte Hernandez}
\author[b,c]{Karl Michael Ziems}
\author[d]{Erik Kjellgren}
\author[d]{Jacob Kongsted}
\author[c]{Sonia Coriani}
\author[{\textdagger},a]{Stephan P. A. Sauer}

\affil[a]{Department of Chemistry, University of Copenhagen, DK-2100 Copenhagen, Denmark.}
\affil[b]{School of Chemistry, University of Southampton, Highfield, Southampton SO17 1BJ, United Kingdom}
\affil[c]{Department of Chemistry, Technical University of Denmark, DK-2800 Kongens Lyngby, Denmark.}
\affil[d]{Department of Physics, Chemistry and Pharmacy, University of Southern Denmark, DK-5230 Odense, Denmark.}

\corres{\textsuperscript{\textasteriskcentered}Corresponding author: \href{mailto:author.one@institute.org}{rehe@chem.ku.dk}.}

\journalname{This research article was typeset in LaTeX with the rho class.}
\journal{København Universitet}
\theday{\today}

\begin{abstract}

Nuclear gradients and Hessians are fundamental quantities in computational chemistry, essential for a wide range of applications including geometry optimization, vibrational spectroscopy, and molecular property calculations. In this work, we present their analytical implementation on quantum hardware. The methodology is formulated within an active-space framework combining orbital optimization and linear-response theory.

On the quantum-computing side, the approach employs the tiled unitary product state (tUPS) ansatz to directly evaluate the tensor elements required for solving the response equations. Moreover, the expectation values are corrected using an adapted confusion-matrix error-mitigation scheme in combination with post-selection criteria. The resulting workflow is assessed on molecular hydrogen and on water through the calculation of potential energy surfaces, nuclear gradients, Hessians, and vibrational frequencies, enabling the evaluation of both its capabilities and current limitations.

The results demonstrate good performance for the hydrogen molecule, whereas the water molecule provides a more demanding test of quantum-hardware resources and highlights the trade-offs associated with error-mitigation strategies. The quantified analysis of the results identify the main sources of errors, suggesting improvement directions for more accurate quantum computer applications. 
\end{abstract}


\begin{document}

    \maketitle
    \thispagestyle{firststyle}


\section{Introduction}

Quantum chemistry simulations on real quantum hardware are proving to be as interesting as they are difficult \cite{MichaelNielsen2000,cao2019quantum,motta2022emerging,paudel2022quantum}. 
Current methods for modeling molecules on quantum processing units (QPU) are built on math and techniques that were first developed for classical computers \cite{Parr1989,Helgaker2013a,Cramer2013,Bloino2016,Loeffelsender2024}.  Yet, they must be fundamentally reconsidered in the context of quantum-classical hybrid algorithms and the limitations of near-term quantum devices.
To realize practical utility of quantum processors in chemistry, the path towards Feynman's dream \cite{Feynman_1982} has required continuous adaptation and reinvention across a wide range of applications: on ground state  \cite{AspuruGuzik2005,Bravyi2017,Grimsley2019,McArdle2020,Tang2021,Cao_2021,Jones_2022,Eddins_2022,Pavosevic2023} and excited state \cite{Ollitrault2020,Ibe_2022,Asthana2023,Kumar2023,jensen2024quantum,zheng2024quantum,ziems2024options,von2026reduced} electronic structure; NMR spectra \cite{seetharam2023digital,ossorio2024simulating,khedri2024impact,fuglsbjerg2026orbital}; hyperfine coupling \cite{jensen2025hyperfine}; vibrational and vibronic structure \cite{Joshi_2014,Huh_2015,Huh_2017,Shen_2018,Sawaya_2019,McArdle_2019,Wang_2020,Wang2020,Ollitrault_2020,Sawaya_2020,Sawaya_2021,Majland2024,Bao2024,OlarteHernandez2024a}; accounting for solvent effect \cite{rossmannek2023quantum,Kjellgren2024, selisko2025dynamical,reinholdt2025self}; and non-linear optical responses \cite{Bruschi_2024}. 

Among the major impediments for reliable quantum calculations on noisy intermediate-scale quantum (NISQ) computers is the accumulation of hardware- and control-induced errors, including gate infidelities, decoherence, and measurement errors, which distort circuit outputs and limit the achievable circuit depth \cite{Devitt_2013,preskill2018quantum, Roffe_2019}. While quantum error correction (QEC) promises fully reliable computation in the long run, it remains unrealizable on current devices due to the enormous overhead in physical qubits and stringent gate fidelity thresholds it requires \cite{shor1996fault, aharonov1997fault}. Quantum error mitigation (QEM) offers a pragmatic near-term alternative: rather than correcting errors at the hardware level, QEM methods reduce their effect through a combination of additional quantum circuit executions and classical post-processing of the resulting measurement statistics \cite{cao2021nisq,suzuki2022quantum,cai2023quantum}. A broad range of QEM techniques have been proposed and studied \cite{khan2024error}, including zero-noise extrapolation \cite{temme2017error, li2017efficient}, probabilistic error cancellation \cite{temme2017error, endo2018practical}, Clifford data regression \cite{czarnik2021error,zhao2025quantum}, readout error mitigation \cite{bravyi2021mitigating}, and Ansatz-based gate and readout error mitigation \cite{ziems2025understanding,rasmussen2025cost}. These methods generally trade an increased sampling cost for improved accuracy, therefore, QEM remains an active and ongoing area of research as it progresses toward useful quantum computations on near-term hardware \cite{cai2023quantum,google2025quantum}.

Molecular energy gradients and Hessians with respect to nuclear coordinates are central quantities for transition state searches and thermodynamic properties \cite{jensen2017introduction,cramer2013essentials}. Moreover, the nuclear gradient, defined as the first derivative of the electronic energy with respect to the nuclear coordinates, is the main ingredient in geometry optimization algorithms and their forces are essential for molecular dynamics simulation. The nuclear Hessian, the second derivatives matrix, is key to accelerating the geometry optimization convergence and retrieving harmonic vibrational frequencies and the associated normal modes \cite{PeterW_Atkins2010}. 
Computing vibrational frequencies represents a first step toward other key nuclear-dependent properties such as force constants, reaction pathways, and vibrational spectroscopy \cite{Helgaker2013a}.
Analytical evaluation of these derivatives is generally preferred over numerical finite-difference approaches, as the latter abruptly scales in computational cost with system size \cite{Helgaker2013}. The theoretical framework for analytical gradients was pioneered by Pulay \cite{pulay1969ab} and later extended to a wide range of electronic structure methods including Hartree-Fock, 
density functional theory 
and coupled cluster theory---
see, e.g., Refs.~\cite{Helgaker1988,Helgaker1992,Hald2023,Feng2019,Schnack-Petersen2022} and references therein.

Classical methods for computing these quantities scale poorly with system size, making quantum simulation an attractive alternative for medium-sized molecules \cite{McArdle2020}. Within hybrid quantum-classical frameworks, analytical methods have been developed for computing first-order energy gradients for ground and excited states \cite{mitarai_theory_2020, parrish_hybrid_2019, parrish_analytical_2021}, nonadiabatic couplings for locating conical intersections \cite{yalouz_analytical_2022, omiya_analytical_2022}, second-order derivatives such as polarizabilities \cite{nakagawa_analytical_2023, obrien_calculating_2019}, and efficient formulations exploiting low-rank factorization for large-scale molecular systems \cite{obrien_efficient_2022, hohenstein_efficient_2023}, as well as geometry optimization algorithms directly on quantum hardware \cite{delgado_variational_2021, azad_quantum_2022}.

This article is part of those continuous efforts to retrieve molecular properties on real quantum hardware. Specifically, the work herein focuses on simulating and retrieving nuclear-dependent properties using the Ansatz-based readout and gate error mitigation technique ($\bm M_0$) \cite{ziems2025understanding}, which has previously been employed to calculate molecular properties in quantum experiments \cite{ziems2025understanding, jensen2025hyperfine, rasmussen2025cost}. The nuclear gradient and nuclear Hessian are calculated using their analytical formulae by measuring the density matrices and orbital response Hessian on a real QPU, with results mitigated via a constructed confusion matrix. The gradients and the vibrational frequencies results are compared to classical full configuration interaction (FCI) \cite{knowles1984new,knowles1989unlimited} and complete active space (CAS) \cite{roos1980complete} references. 
In few words, the novelty of this work lies in the explicit and corrected QPU measurement of tensor elements, that are necessary for calculating the analytical equations of nuclear-dependent properties, with the overall theory embedded within an active space framework. 
To this aim, the use of an adaptable tiled ansatz, the orbital-optimization scheme and assistance of linear response theory, are jointly leveraged as versatile and complementary theoretical tools.

The article proceeds in four parts: it begins with a Theory section (Sec.~\ref{theory}) where energy optimization, quantum circuit, linear response theory, main analytical equations and error mitigation are defined. This is 
followed by the Computational Methods (Sec.~\ref{methods}), defining the quantum circuit, software, and quantum hardware. Subsequently, the Results and Discussions section (Sec.~\ref{results}) illustrates the first and the second nuclear derivative of the energy and the retrieved molecular property. Finally, the Conclusions and Outlook (Sec.~\ref{outlook}) recapitulates on the discussion on the implementation, its limits and future improvements.

\section{Theory}
\label{theory}


This section presents the theoretical framework underlying the methods developed in this work. First, the wavefunction ansatz, Hamiltonian, and energy optimization procedures used in both the classical and quantum approaches are introduced. The tiled quantum circuit is then presented, followed by the linear response formalism required for the implementation of the electronic Hessian. Next, the analytical expressions for the first- and second-order nuclear derivatives of the energy are detailed. Finally, the error-mitigation strategies employed in the quantum experiments are described.

\subsection{Orbital-Optimization Variational Quantum Eigensolver}

Among the first and most promising algorithms for quantum computing is the variational quantum eigensolver (VQE). It is a hybrid quantum-classical algorithm that mixes quantum measurements and classical optimization to minimize a parametrized system. For theoretical chemistry applications, VQE is usually used to obtain the ground state energy and its associated wavefunction. For a molecular system with a geometry $\geo$, the molecular electronic Hamiltonian $\hat H(\geo)$ in second quantization is defined as
\begin{equation}
		\hat H(\geo) =  \sum_{pq} h_{pq}(\geo) \Es_{pq} + \frac{1}{2} \sum_{pqrs} g_{pqrs}(\geo) \Ed_{pqrs} + E_\text{NN}(\geo)\,,
\end{equation}
where $\Es_{pq} =  \adag_{p\alpha}\aan_{q\alpha} + \adag_{p\beta}\aan_{q\beta}$ is the spin adapted singlet single excitation operator, with $\adag$ and $\aan$ the creation and annihilation operators, respectively, for an electron of spin $\{\alpha, \beta\}$ ; the spin adapted singlet double excitation operator $\Ed_{pqrs} =  \Es_{pq} \Es_{rs} - \delta_{qr}\Es_{ps}$; the geometry dependent coefficients $h_{pq}$ and $g_{pqrs}$ are the one- and two-electron integrals, respectively; the last term $E_\text{NN}$ is the nuclear-nuclear repulsion energy. The two sums run over all the spatial orbitals denoted by $p$, $q$, $r$, and $s$. 
For the regular VQE implementation, the parametrized wavefunction (ansatz) is defined via a set of parameters $\bm \Theta$, and a unitary operator $\hat U$, 
as
\begin{equation}
    \ket{\Psi(\geo, \bm \Theta)} = \hat U(\bm \Theta) \ket{\text{CSF}(\geo)} \,,
\end{equation}
where $\ket{\text{CSF}}$ is a single configuration state function reference, which is usually the Hartree-Fock state. The aim is to variationally minimize the expectation value of the Hamiltonian with respect to the variational parameters. This minimization process yields the optimal wavefunction parameters $\bm \Theta_*$, which results in the minimal energy
\begin{equation} \label{eq:E_min_MC}
	\begin{aligned}
		E_{\bm \Theta_*} &= \min_{\bm \Theta} \bra{\Psi(\geo, \bm \Theta)}\hat H(\geo) \ket{\Psi(\geo, \bm \Theta)} \,,
	\end{aligned}
\end{equation}
associated with the parametrized wavefunction.

The VQE procedure can be applied on just a specific subspace of the Hilbert space in order to save computational resources by dramatically reducing the number of required parameters and operations. The latter is referred to as the active space (AS) method, where the Hilbert space is partitioned into inactive, active, and virtual orbitals \cite{siegbahn1980comparison,roos1980complete,siegbahn1981complete}. 
In order to extend the theory and account for intra-spaces optimization, an additional single rotation operator is added to the variational wavefunction 
\begin{equation}
	\ket{\Psi(\geo, \bm \Theta)} 
    = \exp \Big( - \hat \kappa (\bm \kappa) \Big) \hat U(\bm \theta)\ket{\text{CSF} (\geo)} \,.
\end{equation}
where the $\bm \kappa = \{\kappa_{pq}\}$ is the set of orbital parameters, $\bm \theta = \{\theta_{l}\}$ is the configuration (active space) parameters, and thus $\bm \Theta = \{\kappa_{pq}, \theta_{l}\}$ is the variational set spanned by the kappas and the thetas. The $\hat \kappa$ operator is defined as
\begin{equation}
	\begin{aligned}
		\hat \kappa (\bm \kappa) &= \sum_{pq} \kappa_{pq} \left( \Es_{pq} - \Es_{qp}\right) 
        = \sum_{pq} \kappa_{pq} \left( \Es_{pq} - \Es_{pq}^\dagger\right) \,,
	\end{aligned} 
\end{equation}
where the subscripts $p$ and $q$ denote orbital labels associated with different orbital spaces: inactive-to-active ($vi$), inactive-to-virtual ($ai$), active-to-virtual ($av$), and active-to-active ($vw$). In contrast, the subscript $l$ in $\theta_l$ denotes a general index that may refer to different excitation operators, $\hat{\theta}_l$ and $\hat{\theta}_l^\dagger$, depending on the chosen method.
The explicit geometry dependence of the parameters has been omitted to render the notation less cluttered. 

Finally, the orbital-optimization variational quantum eigensolver (oo-VQE) consists of taking advantage of the fact that the orbital rotations can be computed classically, while the most expensive quantum calculation within the active space can be tackled using the quantum computer. Rather than applying the orbital rotation operator directly to the state, the one- and two-electron integrals of the Hamiltonian are transformed using the classical processor, while the circuit parameters $\boldsymbol{\theta}$ are optimized in the usual VQE scheme. In other words, the $\hat \kappa$ operator acts upon the Hamiltonian by solely affecting the values of the integrals
\begin{equation} \label{eq:Hamiltonian_rotated}
	\hat H(\geo, \bm \kappa) = \sum_{pq} h_{pq} (\geo,\bm \kappa)~ \Es_{pq}  + \frac{1}{2}\sum_{pqrs} g_{pqrs}(\geo, \bm \kappa)~ \Ed_{pqrs} \,,
\end{equation}
where
\begin{equation}
	h_{pq} (\geo,\bm \kappa) = \sum_{p'q'} [\exp(\bm \kappa)]_{q'q}  h_{p'q'}(\geo) [\exp(-\bm \kappa)]_{p'p} \,,
\end{equation}
\begin{equation}
	g_{pqrs}(\geo, \bm \kappa) = \sum_{p'q'r's'} [\exp(\bm \kappa)]_{s's} [\exp(\bm \kappa)]_{q'q} g_{p'q'r's'}(\geo) [\exp(-\bm \kappa)]_{p'p}  [\exp(-\bm \kappa)]_{r'r} \,.
\end{equation}
This splitting of the optimization into an active-space part on the quantum hardware and an orbital rotation part on classical hardware is the key advantage of the oo-VQE approach.

\subsection{Unitary Product State Ansatz and Orbital Optimization}

\begin{figure}[H]
\centering
\includegraphics[width=0.7\columnwidth]{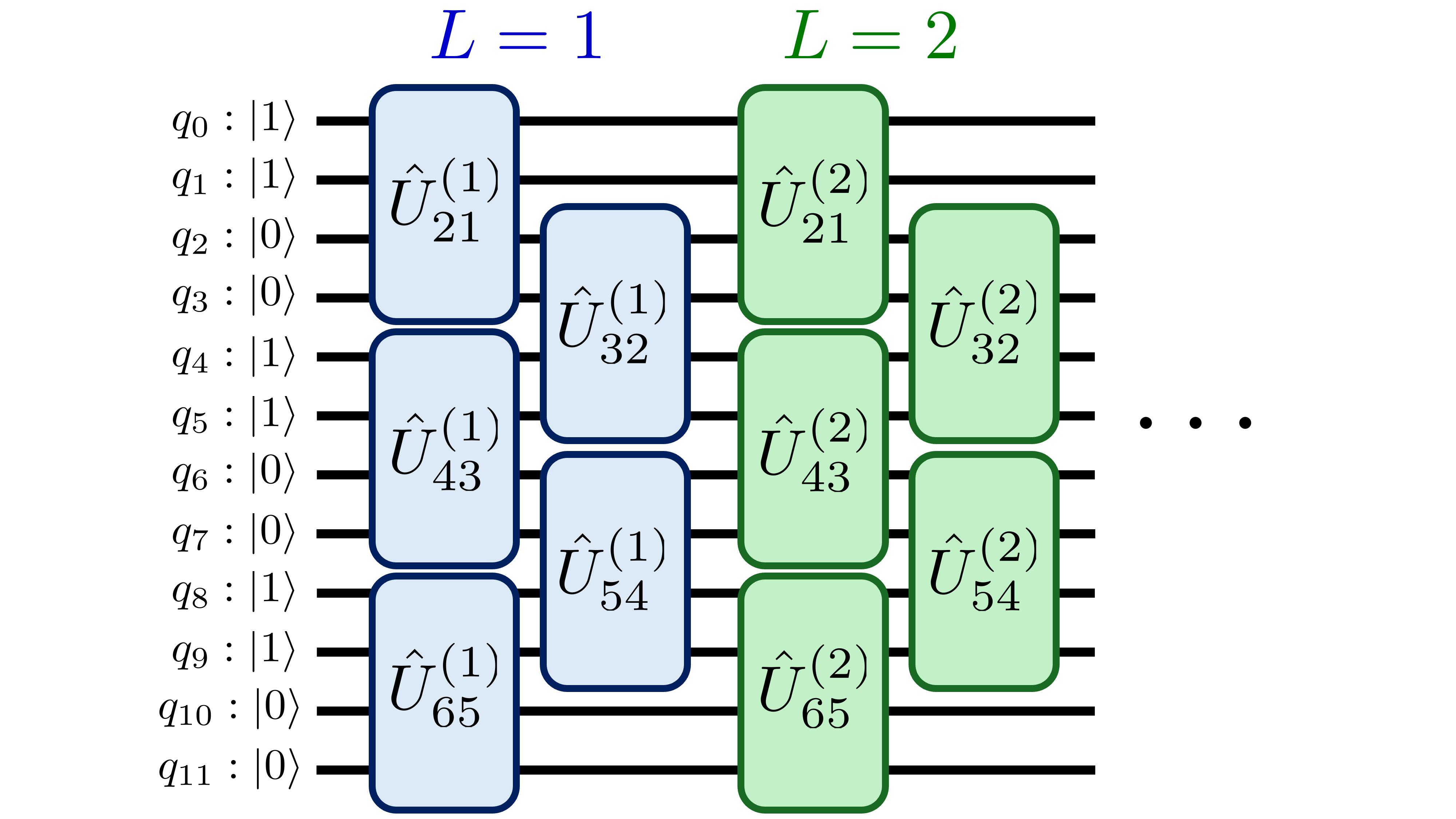}
\caption{The tiled-UPS ansatz with the perfect pairing ordering as proposed by Burton \cite{burton2024}. The first two operator columns correspond to the first layer (blue), and the combination of the first and second columns (blue and green) corresponds to a two-layer ansatz, and so on.}
\label{fig:pptUPS_2}
\end{figure}

For this work, the perfect pairing tiled unitary product state (pp-tUPS) introduced by Burton \cite{burton2024} is chosen as the ansatz. The latter is a tiled-based circuit customizable by its number of layers. The tiles are repetition units of quantum gate operations that link four different spin orbitals at a time. In turn, a layer consists of two well-defined column sets of tiles that relate the whole qubit space, see \figref{fig:pptUPS_2}. Each tile $\hat{U}^{(m)}_{pq}$ takes the form
\begin{equation}
\hat{U}^{(m)}_{pq} 
 = \exp\left(\theta^{(m)}_{pq,1} \hat{\theta}^{(1)}_{pq} \right) \exp\left(\theta^{(m)}_{pq,2} \hat{\theta}^{(2)}_{pq} \right) \exp\left(\theta^{(m)}_{pq,3} \hat{\theta}^{(1)}_{pq} \right) \,,
\end{equation}
where $\hat \theta^{(1)}_{pq} = \left(\hat{E}_{pq}-\hat{E}_{pq}^\dagger\right)$ and $\hat \theta^{(2)}_{pq}=\frac{1}{2}\left(\hat{E}_{pq}^2 - \hat{E}_{pq}^{\dagger\,2}\right)$
are the spin-adapted single- and the pair-double excitation operators acting upon the $p$ and $q$ spatial orbital indices, and $m$ refers to the layer number. 
It must be noted that the pair-double excitation operator differs from its original formulation by the factor 
$\tfrac{1}{2}$.
The tUPS operator for an $n$-qubit system is then
\begin{equation}
\hat{U}_{\mathrm{tUPS}(L)}(\boldsymbol{\theta}) = \prod^L_{m=1}\left(\prod^R_{p=1}\hat{U}^{(m)}_{2p+1,2p}\prod^S_{p=1}\hat{U}^{(m)}_{2p,2p-1}\right)\,,
\end{equation}
where $L$ is the number of layers, and $S =\left\lfloor\frac{n}{4} \right\rfloor$ and $R = \left\lfloor\frac{n-2}{4}\right\rfloor$ are the number of tiles in the first and second column, respectively. The ansatz presents an adaptable layer structure that, by repeating it, yields a wavefunction that approaches the CASSCF one, becoming equivalent in the infinite layer limit. The structure of the ansatz is illustrated in \figref{fig:pptUPS_2}, where the perfect pairing ordering of the tiles is shown.

The pp-tUPS ansatz is embedded within the oo-VQE framework, where the ground-state energy is obtained by variational minimization over both sets of parameters
\begin{equation}
E_\mathrm{UPS} = \min_{\boldsymbol{\theta}, \boldsymbol{\kappa}} \langle \mathrm{UPS}(\boldsymbol{\theta}) | \hat{H}(\geo, \boldsymbol{\kappa}) | \mathrm{UPS}(\boldsymbol{\theta}) \rangle \,,
\end{equation}
where the unitary wavefunction reads
\begin{equation}
\ket{\mathrm{UPS}(\bm \theta)} = \hat{U}_{\mathrm{tUPS}}(\boldsymbol{\theta}) \ket{0011\cdots 0011} \, .
\end{equation}
For the remainder of the article, the ansatz wavefunction is written as 
$\ket{0}$ for shorthand, as it becomes the reference wavefunction.

\subsection{Linear Response}

As a theoretical preamble for the geometry Hessian (\textit{vide infra}), let us define the linear response (LR) method. In general, the LR framework allows for calculation of several properties such as excitation energies, oscillator strengths and rotational strengths, electric-dipole polarisabilities. 
For instance, the generalized eigenvalue equation to retrieve the excitation energies is~\cite{olsen1985linear,helgaker2012recent}
\begin{equation}
    \textbf{E}^{[2]}\boldsymbol{\beta}_k = \omega_k \textbf{S}^{[2]}\boldsymbol{\beta}_k \,, \label{eq:LR_exc}
\end{equation}
where $\textbf{E}^{[2]}$ is the electronic LR Hessian,
$\textbf{S}^{[2]}$ is the metric, 
$\boldsymbol{\beta}_k$ is the excitation vector and $\omega_k$ the corresponding excitation energy for the $k^\text{th}$ excited state. 
The full description of the metric and the excitation vector can be found in SI, as they are not necessary for the remainder of the paper. The important components reside within the LR Hessian which is defined in terms of the required submatrices:
\begin{align}
 \textbf{E}^{[2]} &= \begin{pmatrix}
    {\boldsymbol{A}} & {\boldsymbol{B}} \\
      {\boldsymbol{B}}^* & {\boldsymbol{A}}^*           
     \end{pmatrix} \,.
\end{align}
The $\boldsymbol{A}$ and $\boldsymbol{B}$ submatrices involve double commutators between the Hamiltonian operator $\hat{H}$, the orbital rotation operators $\hat{q}_{\mu}$ and the active space excitation operators $\hat{G}_m$ :
\begin{align}
    \bm{A} & ~=~\begin{pmatrix}
\left<0\left|\left[\hat{q}_\mu^\dagger,\hat{H},\hat{q}_{\nu}\right]\right|0\right>
& \left<0\left|\left[ \hat{q}^\dagger_\mu ,\hat{H}, \hat{G}_{m}\right]\right|0\right>  \\
\left<0\left|\left[\hat{G}_{n}^\dagger,\hat{H},\hat{q}_{\nu}\right]\right|0\right>
& \left<0\left|\left[\hat{G}_{n}^\dagger,\hat{H},\hat{G}_{m}\right]\right|0\right>
\end{pmatrix} \,, \\
    \boldsymbol{B} & ~= \begin{pmatrix}
\left<0\left|\left[\hat{q}_\mu^\dagger,\hat{H},\hat{q}_{\nu}^\dagger\right]\right|0\right>
& \left<0\left|\left[\hat{q}^\dagger_\mu ,\hat{H}, \hat{G}^\dagger_{m} \right]\right|0\right> \\
\left<0\left|\left[\hat{G}_{n}^\dagger,\hat{H},\hat{q}_{\nu}^\dagger\right]\right|0\right>
& \left<0\left|\left[\hat{G}_{n}^\dagger,\hat{H},\hat{G}_{m}^\dagger\right]\right|0\right>
\end{pmatrix} \,,
\end{align}
where the symmetrized double commutator was used
\begin{equation}
    \forall \{ \hat O , \hat P\} \in \{ \hat q,\hat G \},\hspace{0.5cm} [\hat O, \hat H, \hat P] = \frac{1}{2}\left( \left[ \hat O, [\hat H, \hat P ] \right] +  \left[ \hat P, [\hat H, \hat O ] \right] \right) \,.
\end{equation}
The submatrices satisfy the following properties:
\begin{equation} \label{eq:AB_properties}
    \boldsymbol{A} = \boldsymbol{A}^\dagger \hspace{0.4cm},\hspace{0.4cm}  \boldsymbol{B} = \boldsymbol{B}^\text{T} \,.
\end{equation}
The choice of operators $\{\hat q, \hat G \}$ is not trivial, so as a first approach the naive operators are used \cite{ziems2024options,kjellgren2024divergences}. For our purposes, the use of naive operators for the full space calculation of H$_2$ is exact, whereas for the H$_2$O active-space calculation it is an approximation \cite{kjellgren2025redundant}.
The naive operators are truncated to spin-adapted singles and doubles excitations as~\cite{paldus1977application,piecuch1989orthogonally},
\begin{align}
    \hat q_{pq} = \frac{1}{\sqrt{2}}\hat{E}_{pq}~, \quad\textrm{with} \quad  pq=\{vi, ai,av \} \equiv \mu  
    \label{eq:q}
\end{align}
and
\begin{align}
    \hat{G} \in \Bigg\{\frac{1}{\sqrt{2}}\hat{E}_{v_av_i},\quad &\frac{1}{2\sqrt{\left(1+\delta_{v_av_b}\right)\left(1+\delta_{v_iv_j}\right)}}\left(\hat{E}_{v_av_i}\hat{E}_{v_bv_j} + \hat{E}_{v_av_j}\hat{E}_{v_b v_i}\right), \label{eq:G} \\\nonumber
    &\frac{1}{2\sqrt{3}}\left(\hat{E}_{v_av_i}\hat{E}_{v_bv_j} - \hat{E}_{v_av_j}\hat{E}_{v_bv_i}\right)\Bigg\}.
\end{align} 
Similarly to the oo-VQE framework, the orbital rotation operators $\hat q_{pq}$ accept indices $pq$ between the inactive to active ($vi$), inactive to virtual ($ai$), and active to virtual ($av$) spaces, whereas the AS orbital excitation operators $\hat{G}_m$ are confined within the AS.
The AS indices $v$ carry subscripts $\{a,b\}$  and $\{i,j\}$ to indicate orbitals that are virtual (unoccupied) and inactive (occupied), respectively, in the reference determinant.

\subsection{Analytical Nuclear Gradient and Hessian}

The main equation for the expression of the energy depending on the nuclear coordinates $\geo$ and the electronic parameters of the wavefunction $\bm \Theta$ was derived by Helgaker and Jørgensen \cite{Helgaker1986}. As a summary, their derivation started by expanding the energy with respect to the variational parameters $\bm \Theta$ ($\lambda$ in their notation) of the wavefunction on the unperturbed case. Then, both the parameters and their associated expansion coefficients were also expanded in Taylor series now with respect to the geometry perturbation. Finally, 
the energy is purely expanded with respect to the perturbation and, by subsequent identification between the different orders of the perturbation and using the stationary parameters, the final expression of the energy expansion can be simply written as
\begin{equation}
    E(\geo,\bm\Theta) = 	E^{(0)} + E^{(1)} \geo +  \frac{1}{2} E^{(2)} \geo^2 + ...  \,,
\end{equation}
where the expansion coefficients $E^{(n)}$ ($\varepsilon^{(n)}$ in their notation) depend in turn on the different orders of the variational parameters.
In Helgaker and Jørgensen's article, after imposing some stability conditions on the variational parameters, equation grouping and variable identification, the different terms in the expansion can be explicitly and analytically expressed. These expressions are the ones used for the following.

\subsubsection{Nuclear Gradient}

Helgaker and Jørgensen \cite{Helgaker1986,Helgaker1988,Helgaker1992} expressed the first derivative of the energy with respect to the nuclear coordinates as 
\begin{equation} \label{eq:explicit_gradiet}
	\begin{aligned}
	E^{(1)}_{A \alpha}  &=  \text{Tr} \left( D_\text{AO} h_{\text{AO}~A\alpha}^{(1)} \right)  + \frac{1}{2} \text{Tr} \left(d_\text{AO} g_{\text{AO}~A\alpha}^{(1)} \right)  - \text{Tr}\left( S_{\text{AO}~ A\alpha }^{(1)} F_\text{AO} ^{[0]} \right) + E^{(1)}_{\text{NN}~A\alpha}  \\	
	&=  \sum_{\mu\nu} D_{\text{AO}~\mu\nu} h_{\text{AO}~A\alpha , \nu\mu}^{(1)} 
	+ \frac{1}{2} \sum_{\mu \nu \lambda \sigma} d_{\text{AO}~\mu \nu \lambda \sigma} g_{\text{AO}~A\alpha, \mu \nu \lambda \sigma}^{(1)} 
	 - \sum_{\mu \nu} S_{\text{AO}~ A\alpha,  \mu \nu  }^{(1)} F_{\text{AO}~\nu \mu  } ^{[0]} + E^{(1)}_{\text{NN}~A\alpha} \,,
	\end{aligned} 
\end{equation}
The nuclear gradient is a vector that contains the information of how much the energy changes considering a particular atom $A$, following a specific Cartesian direction $\alpha \in \{x, y, z\}$.
In Eq.~\eqref{eq:explicit_gradiet}, $D_\text{AO}$ and $d_\text{AO}$ are the one-electron and two-electron reduced density matrices (1-RDM and 2-RDM), respectively; $F_\text{AO} ^{[0]}$ is the effective density, also called the generalized Fock matrix; $h_\text{AO}^{(1)}$, $g_\text{AO}^{(1)}$, and $S_\text{AO}^{(1)}$ are, respectively, the first derivatives of the one-electron, two-electron, and overlap integrals with respect to the nuclear coordinates; $E^{(1)}_\text{NN}$ is the first derivative of the nuclear-nuclear energy. 
The lower-script `AO', indicates that the tensors are represented in the AO basis. 
The density matrices are either a two-rank tensor, or a four-rank tensor. The other lower-scripts represent the coordinate index that has been split as two indices $A\alpha$, and the AO indices as $\mu$, $\nu$, $\lambda$, and $\sigma$.  The formulae for constructing all these elements are reported in the Supplementary Information (SI). 
The RDM elements can be both explicitly expressed as expectation values of the single and the double excitation operators, and, as such, they are the elements to be measured on the quantum computer:
\begin{equation}
    D_{\text{AO}\,\mu\nu} = \sum_{pq} C_{\mu p}C_{\nu q} \bra{0} \Es_{pq} \ket{0},
\end{equation}
\begin{equation}
    d_{\text{AO}\,\mu\nu\lambda\sigma} = \sum_{pqrs}C_{\mu p}C_{\nu q}C_{\lambda r}C_{\sigma s}\bra{0} \Ed_{pqrs} \ket{0}.
\end{equation}
where $\boldsymbol{C}$ is the matrix of molecular orbital coefficients.

The first two traces in Eq.~\eqref{eq:explicit_gradiet} represent the Hellmann-Feynman terms, which are the \textit{`immediate'} derivatives of the energy.
The third trace is the Pulay forces or the connection correction. 
The Pulay contribution corrects for the fact that the variationally optimized orbitals change when the nuclei move. These Pulay forces disappear in the case of i) a complete basis, or ii) the basis is independent of nuclear positions (e.g., plane waves), or iii) in the limit of exact wavefunctions. This is the same expression as the one used in Sugisaki \textit{et. al} \cite{sugisaki2022quantum}, with the difference that their integrals are the connection corrected version expressed in the MO basis. Within a quantum computer framework, working in the MO basis is more natural. 
Herein, we chose to work in the AO basis, as these are the basis sets commonly implemented in quantum chemistry classical programs, and foremost it allows us to explicitly follow and quantify the effects of the change of basis of a moving geometry.
As a representative of the gradient's magnitude (or modulus), let us define the gradient force
\begin{equation} \label{eq:Grad_force}
	\parallel E^{(1)} \parallel = \pm \sqrt{ \sum_{A\alpha} {E^{(1)}_{A \alpha}}^2} \,.
\end{equation}
This is the expression used when plotting the gradient, instead of the full 6-value gradient vector. The plus-minus sign indicates that the overall gradient sign is considered and applied \textit{a posteriori}, representative of the bond shortening or stretching behavior induced by the force.  

\subsubsection{Nuclear Hessian}

Helgaker and Jørgensen also derived the second derivative of the energy with respect to the nuclear coordinates \cite{Helgaker1986,Helgaker1988,Helgaker1992}. 
Let us define the nuclear Hessian as the contribution of three different terms 
\begin{equation} \label{eq:NuclearHessian_complet}
    E^{(2)} = E^{(2)}_\text{static} + E^{(2)}_\text{relaxed} + E^{(2)}_\text{NN} \,,
\end{equation}
where the static (or unrelaxed) tensor elements can be explicitly expressed for a pair of atoms $A$ and $B$, in the $\alpha$ and $\beta$ directions:
\begin{equation} \label{eq:static_Hessian}
	\begin{aligned}
		E^{(2)}_{\text{static}~ A \alpha B\beta}  &=  \text{Tr} \left( D_\text{AO} h_{\text{AO}~A\alpha B\beta}^{(2)} \right)  + \frac{1}{2} \text{Tr} \left(d_\text{AO} g_{\text{AO}~A\alpha B\beta}^{(2)} \right)  - \text{Tr}\left( S_{\text{AO}~ A\alpha B\beta}^{(2)} F_\text{AO} ^{[0]} \right)   \\	
		& ~~~~ -2 \text{Tr}\left( S_{A\alpha}^{(1)}F_{B\beta}^{[1]} \right)  + \text{Tr} \left(S_{A\alpha}^{(1)} S_{B\beta}^{(1)} F^{[0]}  \right) + \frac{1}{2} \text{Tr} \left[   S_{A\alpha}^{(1)} (S^{(1)},F^{[0]})_{B\beta}\right] \\ 
		{}\\
		&=  \sum_{\mu, \nu} D_{\text{AO}~\mu\nu} h_{\text{AO}~A\alpha B\beta , \nu\mu}^{(2)} 
		+ \frac{1}{2} \sum_{\mu \nu \lambda \sigma} d_{\text{AO}~\mu \nu \lambda \sigma} g_{\text{AO}~A\alpha B\beta, \mu \nu \lambda \sigma}^{(2)} 
		- \sum_{\mu \nu} S_{\text{AO}~ A\alpha B\beta,  \mu \nu  }^{(2)} F_{\text{AO}~\nu \mu  } ^{[0]}  \\
		& ~~~~ -2 \sum_{p q} S_{A\alpha,  pq  }^{(1)} F_{B\beta,  qp  } ^{[1]}
		+\sum_{p q r} S_{A\alpha,  pq  }^{(1)} S_{B\beta,  qr }^{(1)}  F_{ rp  } ^{[0]}
		+ \frac{1}{2} \sum_{pq} S^{(1)}_{A\alpha, pq} 	(S^{(1)}, F^{[0]})_{B\beta, qp} \,,
		\end{aligned}
	\end{equation}
where now the second derivatives of the different integrals with respect to the nuclear coordinates appear. 
Within the additional new three traces, the absence of lower-script indicates that the tensors are expressed in the MO basis: the first derivative of the overlap $S^{(1)}$, as well as the zeroth and first derivative of the effective (Fock) density, $F^{[0]}$ and $F^{[1]}$, respectively, and the metric correction $(S^{(1)},F^{[0]})$. 
The latter corresponds to a corrected Fock matrix, constructed from integrals that are one‑index transformed by $S^{(1)}$ (see SI for its explicit equation) which accounts for the change in the atomic orbital basis with nuclear displacement.
The first two traces are the Hellmann-Feynman(-like) Hessians, whereas the remaining traces are the Pulay-like connection correction contributions.

The second term is referred to as the relaxed Hessian $E_\text{relaxed}^{(2)}$, also called the coupled perturbed (CP) response contribution. It encodes the orbital response to nuclear displacement and approximately vanishes for frozen orbitals. 
It is composed of the first-order property vector (or driving term) $f^{(1)}$, and the (orbital) response vector  $\lambda^{(1)}$,
\begin{equation}
E_\text{relaxed}^{(2)} = f^{(1)} \lambda^{(1)} \,.
\end{equation}
These two tensors are related through the CP-MCSCF equation,
\begin{equation} \label{eq:CP_MCSCF}
	\begin{aligned}
	 \underbrace{	\frac{\partial^2 E}{\partial \geo \partial \bm \Theta} }  + 	\underbrace{ \frac{\partial^2 E}{\partial \bm\Theta \partial \bm\Theta} } \underbrace{\frac{\partial\bm\Theta}{\partial \geo} } &= 0 \\
		\hspace{-5cm}	\Leftrightarrow	 \hspace{1cm}  f^{(1)}   \hspace{0.2cm}  +  \hspace{0.2cm} \G^{(0)} \hspace{0.4cm} \lambda^{(1)}  &= 0 \,,
	\end{aligned}
\end{equation}
 which solves for the response vector $\lambda^{(1)}$. The $\G^{(0)}$ tensor is the electronic orbital Hessian, which encodes the double derivative of the energy with respect to the wavefunction parameters. 
Different techniques have been developed to solve Eq.\eqref{eq:CP_MCSCF} \cite{Helgaker1986}, but due to the small size of the systems, the equation is here solved by explicit full-space measurement and inversion of the electronic Hessian. The response term admits two expressions: one for the $\kappa$-orbital parameters, and another for the $\theta$-configuration rotations
 \begin{equation}
 	f^{(1)} = \frac{\partial^2 E}{\partial \geo \partial \bm\Theta} = \begin{pmatrix}
 		f^{(1)}_{\bm \kappa}  \\ f^{(1)}_{\bm \theta} 
 	\end{pmatrix}
 \end{equation}
 Instead of explicitly writing the orbitals to which the $\Theta$-operators acts upon, the set of operators is just numbered with a $n$ cardinal, $\bm \Theta  = \{\kappa_n\}_{n \in [1, N_\kappa]} \cup \{\theta_n\}_{n \in [N_\kappa+1, N_\kappa+ N_\theta]}$, where $N_\kappa$ and $N_\theta$ are the number of $\kappa$ and $\theta$ parameters, respectively. With this notation, the $n^\text{th}$ element of the driving vector is
\begin{equation}
f^{(1)}_{\Theta_n~A\alpha} = \left(  f^{(1)}_\Theta  \right)_{A\alpha,~n}
	= \frac{\partial}{\partial R_{A\alpha}}\!\left[\frac{\partial E}{\partial \Theta_n}\right]
    =
\begin{cases}
\langle0|\,[\Es_{n} - \Es_{n}^\dagger,\,\hat{H}^{(1)}_{\text{OMO}~A\alpha}]\,|0\rangle , &  n \in [1,N_\kappa] \\[10pt]
\langle0|\,[\hat \theta_n - \hat \theta_n^\dagger,\,\hat{H}^{(1)}_{\text{OMO}~A\alpha}]\,|0\rangle , &  n \in [N_\kappa+1,N_\kappa+N_\theta]
\end{cases}
\end{equation}
where $\ket{0} = \ket{\mathrm{UPS}(\bm \theta)}$ is the reference wavefunction, and $\hat{H}^{(1)}_{\text{OMO}~A\alpha}$ is the first-order derivative Hamiltonian built from $h^{(1)}_{\text{OMO}~A\alpha}$ and $g^{(1)}_{\text{OMO}~A\alpha}$, expressed in the orthogonal molecular orbital (OMO) basis (Eq.~\eqref{eq:h-OMO} and Eq.~\eqref{eq:g-OMO} in SI).
Likewise, the orbital Hessian can be compartmentalized by operator-types:
 \begin{equation}
	\G^{(0)} = \frac{\partial^2 E}{\partial\bm\Theta \partial\bm\Theta} = \begin{pmatrix}
		{}^{\kappa\kappa}\G^{(0)} & {}^{\kappa \theta}\G^{(0)} \\  {}^{\theta\kappa}\G^{(0)}   &   {}^{\theta\theta}\G^{(0)} 
	\end{pmatrix}  \,,
\end{equation}
 where the left upper-scripts denote which pair of orbital-configurational rotation operators are used. The internal structure of the Hessian is reminiscent of the structure of the LR matrices $\bm A$ and $\bm B$. By considering two operators $\hat \Theta_n = \hat O_n - \hat O_n^\dagger$ and $ \hat \Theta_m = \hat O_m - \hat O_m^\dagger$, where the $\hat O = \{\hat q , \hat G\}$ operator defines either a single or double excitation; and redefining the variational parameters $\Theta \leftarrow \Theta - \Theta^{(0)}$, where $\Theta^{(0)}$ is the optimized set of variational parameters of the ground state wavefunction; it is shown that
 \begin{equation}
 \begin{aligned}
	\G^{(0)}_{nm}  &= \frac{\partial^2 E}{\partial \Theta_n\, \partial \Theta_m}\bigg|_0 \\
	&=  \bra{0}  \bigl[\hat{\Theta}_n,\, [\hat{\Theta}_m , \, \hat{H} ]\bigr]  \ket{0} \\
    &= \bra{0}  \bigl[\hat{O}_n - \hat{O}_n ^\dagger,\, [ \hat{O}_m - \hat{O}_m^\dagger , \, \hat{H}]\bigr]  \ket{0} \\
    &= \bra{0}  \bigl[\hat{O}_n - \hat{O}_n ^\dagger,\, [ \hat{O}_m, \hat{H}] - [\hat{O}_m^\dagger , \, \hat{H}]]\bigr]  \ket{0} \\
    &= \bra{0}  \bigl[\hat{O}_n - \hat{O}_n ^\dagger,\, [ \hat{O}_m, \hat{H}] \bigr]  \ket{0} 
    - \bra{0}  \bigl[\hat{O}_n - \hat{O}_n ^\dagger,\, [\hat{O}_m^\dagger , \, \hat{H}]\bigr]  \ket{0} \\
    &= \bra{0}  \bigl[\hat{O}_n ,\, [ \hat{O}_m, \hat{H}] \bigr]  \ket{0} 
    - \bra{0}  \bigl[ \hat{O}_n ^\dagger,\, [ \hat{O}_m, \hat{H}] \bigr]  \ket{0} 
    - \bra{0}  \bigl[\hat{O}_n ,\, [\hat{O}_m^\dagger , \, \hat{H}]\bigr]  \ket{0}
    + \bra{0}  \bigl[ \hat{O}_n ^\dagger,\, [\hat{O}_m^\dagger , \, \hat{H}]\bigr]  \ket{0} \\
    &= - \bra{0}  \bigl[\hat{O}_n ,\, [\hat{H}, \, \hat{O}_m] \bigr]  \ket{0}
       + \bra{0}  \bigl[ \hat{O}_n ^\dagger,\, [\hat{H}, \hat{O}_m ] \bigr]  \ket{0} 
       + \bra{0}  \bigl[\hat{O}_n ,\, [ \hat{H}, \,\hat{O}_m^\dagger]\bigr]  \ket{0} 
       - \bra{0}  \bigl[ \hat{O}_n ^\dagger,\, [\hat{H}, \,\hat{O}_m^\dagger  ]\bigr]  \ket{0} \,,
 \end{aligned}
\end{equation}
where the notation $\big|_0$ indicates that the derivatives are taken in $\bm \Theta = \bm 0$ which, because of the change of variable notation, it is equivalent to do the derivative around the (previous) optimized $\Theta$ parameters. Then, by symmetry of the Hessian:
\begin{equation}
 \begin{aligned}
	2 \G^{(0)}_{nm}  &= \G^{(0)}_{nm} + \G^{(0)}_{mn} \\
    &=  \bra{0}  \left( \bigl[ \hat{O}_n ^\dagger,\, [\hat{H}, \hat{O}_m ] \bigr]  +  \bigl[\hat{O}_m ,\, [ \hat{H}, \,\hat{O}_n^\dagger]\bigr] \right) \ket{0} 
      - \bra{0} \left(\bigl[\hat{O}_n ^\dagger,\,[\hat{H},\,\hat{O}_m^\dagger]\bigr] + \bigl[\hat{O}_m ^\dagger,\,[\hat{H},\,\hat{O}_n^\dagger]\bigr]\right) \ket{0}  \\
     &~ - \bra{0} \left( \bigl[\hat{O}_n ,\, [\hat{H}, \, \hat{O}_m] \bigr] +  \bigl[\hat{O}_m ,\, [\hat{H}, \, \hat{O}_n] \bigr] \right)  \ket{0} + \bra{0} \left( \bigl[\hat{O}_n ,\, [ \hat{H}, \,\hat{O}_m^\dagger]\bigr]  + \bigl[ \hat{O}_m ^\dagger,\, [\hat{H}, \hat{O}_n ] \bigr] \right) \ket{0} \\
     \Longleftrightarrow ~~ \G^{(0)}_{nm} &= \boldsymbol{A}_{nm} - \boldsymbol{B}_{nm} 
     -\boldsymbol{B}_{nm}^* + \boldsymbol{A}_{nm}^*  \,.
 \end{aligned}
\end{equation}
Using the properties of the submatrices in Eq.~\eqref{eq:AB_properties}, and that both submatrices are real:
\begin{equation}
 \begin{aligned}
	 \G^{(0)}_{nm} &= \boldsymbol{A}_{nm} + \boldsymbol{A}_{nm}^\text{T}  - \boldsymbol{B}_{nm} -\boldsymbol{B}_{nm}^\text{T} \\
     &= 2 \left( \boldsymbol{A}_{nm} - \boldsymbol{B}_{nm} \right) \,,
 \end{aligned}
\end{equation}
thus the use of the LR matrices. 
 Finally, the response vector is built upon inversion of the orbital Hessian and product with the driving term
\begin{equation}
	\lambda^{(1)}_{\Theta_n~B\beta} = \left(   \lambda^{(1)}_\Theta  \right)_{B\beta,~n} 
	=  \sum_{m}  \left[ -\G^{(0)}\right]^{-1}_{n,m}   f^{(1)}_{\Theta_m~B\beta} \,,
\end{equation}
and so the final expression of the nuclear relaxed Hessian reads
\begin{equation} \label{eq:relaxed_Hessian}
	\begin{aligned}
		E^{(2)}_{\text{relaxed}~ A \alpha B\beta}  &= \sum_n f^{(1)}_{\Theta_n~A\alpha}\;\lambda^{(1)}_{\Theta_n~B\beta}, \\
		&=  - \sum_n  \sum_{m} f^{(1)}_{\Theta_n~A\alpha}\;  \left[ {}^{\Theta\Theta}\G^{(0)}\right]^{-1}_{n,m}   f^{(1)}_{\Theta_m~B\beta} \,.
	\end{aligned}
\end{equation}

\subsection{Error Mitigation} \label{subsection:ErrorMitigation}

\subsubsection{Ansatz-based readout and gate error mitigation}

The M0 error mitigation method used herein was proposed by Ziems et al.\ \cite{ziems2025understanding} and has since been applied both in simulation and on real quantum hardware \cite{ziems2025understanding,jensen2025hyperfine,reinholdt2025self,rasmussen2025cost}. By definition, one element of the M0 confusion matrix is defined as the transition probability between two states:
\begin{equation}
	\bm M_{0, yx} = \mathbb{P}(\ket y \vert \ket x)
\end{equation}
which reads as the probability of measuring $\ket y$ given that $\ket x$ was prepared as the initial state. In a standard REM confusion matrix, the initial state is prepared using only $X$ gates (\textit{e.g.}\ for a two-qubit system $\ket{x} \in \{ \ket{00},\ X_0 \ket{00} = \ket{10},\ X_1 \ket{00} = \ket{01},\ X_0X_1\ket{00} = \ket{11} \}$). In the M0 approach, the chosen ansatz is additionally applied with all ansatz parameters set to zero:
\begin{equation}
	\ket{x_0} = \hat U( \bm 0) \ket{x}
\end{equation}
This is the state that is repeatedly measured. The statistical measure estimates the frequency of appearance of the different states. For a finite number of shots $N \in \mathbb{N}$, yielding outcomes $\{ \ket{s_i} \}_{i \in [1,N]}$, the frequency estimate for a state $\ket y$ reads
\begin{equation}
	\bm M_{0, yx} = \frac{1}{N} \sum_{i= 1}^N \braket{s_i}{y}
\end{equation}
which uses the orthonormality of the states. 
The resulting matrix captures how much the measurement outcome deviates from the ideal while considering the gate noise drifting.  These transition probabilities are then applied to the measurement results of the actual circuit of interest. Denoting the noisy measurement outcomes $\{ \ket{ r_i } \}_{i \in [1,N]}$, the mitigated probability vector is defined as
\begin{equation}
	\begin{aligned}
		\ket{ \mathbb{P}_\text{mitigated}} &=  	\bm M_{0}^{-1}   \underbrace{\frac{1}{N} \sum_{i= 1}^N \ket{r_i} } \\
		&= \bm M_{0}^{-1} \hspace{0.25cm} \ket{\mathbb{P}_\text{raw}}
	\end{aligned}
\end{equation}
where the raw noisy measurements are algebraically grouped. Finally the mitigated expectation value of an operator $\hat O$, with associated $\{ O_x\}$ eigenvalues and $\{ \ket x \}$ eigenvectors, is 
\begin{equation}
	\left< \hat O \right> = \sum_{x \in \{0,1\}^{n_q}} O_x \bra{x}  \bm M_{0}^{-1} \ket{\mathbb{P}_\text{raw}}
\end{equation}
where $n_q$ is the number of qubits. Measuring the confusion matrix requires a large number of preliminary measurements that scale exponentially with the number of qubits. Limiting its application to small systems, and such, convenient to the case herein. 

\subsubsection{Post-Selection}

As an additional mitigation strategy, after the $\bm M_0$ error mitigation, the Post-Selection method \cite{jensen2025hyperfine} is employed for the diagonal elements of the 1-RDM. It consists of conserving only those bit-strings that respect the expected number of $\alpha$ and $\beta$ electrons separately. This is possible for diagonal elements because their measurement operators commute with the Pauli-Z string and do not change the particle number, conversely to off-diagonal elements involving Pauli-X and -Y operators which might affect it. Discarding these faulty states effectively filters out measurement noise and ensures the 1-RDM trace condition is satisfied.

\section{Computational Methods}
\label{methods}

\subsection{The systems, programs, and computer}

The systems of choice are the hydrogen and the water molecules. For the hydrogen molecule, a full active space AS(2,2) in the STO-3G \cite{Hehre_1969} minimal basis is used, making the wavefunction equivalent to an FCI calculation. The ground state geometry was determined by an FCI geometry optimization using the PySCF program \cite{sun2007python,sun2018pyscf,sun2020recent}, yielding an equilibrium bond length of $d_\text{HH,eq} = 0.7349 \text{ \AA}$. For the water molecule, the geometry was optimized using the same program and basis set, but with an active space of four electrons and four orbitals, AS(4,4). The CASSCF optimization yields an equilibrium oxygen-hydrogen bond length $d_\text{OH,eq} = 1.0279 \text{ \AA}$, with an equilibrium angle $\widehat{HOH} = 96.73^\circ$. 

The energy and gradient references are computed at the CASSCF level of theory using the PySCF program, which integrates routines for both quantities. However, PySCF does not include a routine for computing the Hessian at the CASSCF level, as only the CPHF equations are implemented instead of the necessary CP-MCSCF equations. Therefore, the reference nuclear Hessian was obtained by the finite difference (FD)  method implemented using the central difference method with a geometrical step of $0.001$ \AA; for the electronic Hessian the step is of $0.001$ radians. As further confirmation, the Dalton program \cite{aidas2014d} was used to compare specific gradient and Hessian values, whose validity is mentioned when used in the results section. Moreover, the necessary integrals are obtained or post-constructed using PySCF.

The simulations are carried out using the quantum chemistry software SlowQuant \cite{SlowQuant}. The (pp-)tUPS ansatz and Hamiltonian are mapped using the Jordan-Wigner mapping. Additionally, SlowQuant orchestrates the translation to Qiskit \cite{Qiskit}, the native language of the IBM hardware and programs, launching the QPU jobs in the IBM Pittsburgh backend. The specifics on the Pittsburgh hardware can be seen in \tabref{tab:ibm_pittsburgh}, and notably the individual qubit readout errors are summarized in SI. 
\begin{table}[H] 
	\centering
	\caption{IBM's Pittsburgh hardware overall performance. \label{tab:ibm_pittsburgh}}
	\begin{tabular}{cccccc}
		\toprule
		\textbf{\makecell{Number of\\ qubits}} & \textbf{Couplers} & \textbf{\makecell{Median 2 Qubit\\ error}} & \textbf{\makecell{Mean 2 Qubit error\\ (layered)}} & \textbf{\makecell{2 Qubit error for\\ 4 qubits (layered)} } &  \textbf{\makecell{Processor\\ type}} \\ \addlinespace[0.3em]
		156 & 176 & $1.72\times10^{-3}$ & $ 2.37\times 10^{-3}$ & $1.68 \times 10^{-3}$ &  Heron r3 \\ 
		\bottomrule
	\end{tabular}
	\tabletext{Note: Values taken on the 12$^\text{th}$ May 2026. These values are sensitive to calibration and may vary. }
\end{table}
Moreover, after circuit transpilation and qubit assignment, if a measurement uses an under-performing qubit (with relatively large errors) the result is discarded. 

\subsection{One- and two-layer pp-tUPS ansätze}

  For the first case, the H$_2$/STO-3G system requires only 4 spin-orbitals, hence, 4 qubits. Given its structure, it is expected (and further proved) that a one-layer (pp-)tUPS circuit is enough to be tantamount to an FCI wavefunction. Therefore, the ansatz used for the hydrogen molecule is the one-layer tUPS where the unitary operator is
\begin{equation}
	\hat U_\text{tUPS(1)}(\theta_{21,2}, \theta_{21,3}) = \hat U^{(1)}_{21} = \left(\mathrm{e}^{\theta_{21,1} \left(\hat E_{21} - \hat E_{21}^\dagger\right)} \right) \mathrm{e}^{  \frac{\theta_{21,2}}{2}\left(\hat E_{21}^2 - \hat E_{21}^{\dagger\,2}\right)}  \mathrm{e}^{\theta_{21,3} \left(\hat E_{21} - \hat E_{21}^\dagger\right)} 
\end{equation}
The first exponential between brackets is part of the complete original formulation in Burton's paper \cite{burton2024}. Nonetheless, given the form of the UPS wavefunction, this last (in application) singlet rotation is redundant with respect to the $\kappa$-optimization of the wavefunction in the oo-VQE scheme. Thus, this single excitation rotation is dropped, and only the double excitation and final single excitation operators are considered. 
For the water molecule, the AS(4,4) requires thus 8 spin-orbitals or qubits. It will be demonstrated that a two-layer tUPS wavefunction is necessary to imitate the PySCF-CASSCF(4,4) wavefunction. The two layer operator reads:
\begin{equation}
	\hat U_\text{tUPS(2)}(\bm \theta) =\left( \hat U^{(2)}_{32} \hat U^{(2)}_{43} \hat U^{(2)}_{21} \right) \left( \hat U^{(1)}_{32} \hat U^{(1)}_{43} \hat U^{(1)}_{21} \right)
\end{equation}
Just as before, the final redundant singlet excitation in $\hat U^{(2)}_{32}$ is not considered. After the ansätze application, the resulting wavefunctions are
\begin{equation}
\begin{aligned}
	\ket{\text{tUPS(1)}(\bm \theta)} &= 	\hat U_\text{tUPS(1)} (\bm \theta) \ket{0011} , \\
    \ket{\text{pp-tUPS(2)}(\bm \theta)} &= 	\hat U_\text{tUPS(2)} (\bm \theta)\ket{00110011}
\end{aligned}
\end{equation}
for the hydrogen and the water molecule, respectively. 
The circuits summary is in \tabref{tab:pptUPS_ansatz_details}, and their visualization, in their ideal and transpiled forms, are presented in \figref{fig:pptUPS_circuit} and \figref{fig:pptUPS_circuit_transpiled} in SI.  
\begin{table}[H] 
	\centering
	\caption{Details on the two different ansätze for the hydrogen and water molecules. The entanglers are the CNOT and the CZ gates for the ideal and the transpiled circuits, respectively. \label{tab:pptUPS_ansatz_details}}
	\begin{tabular}{cc|lcc}
		\toprule
		\textbf{Molecule} & \textbf{Circuit level} & & \textbf{Ideal circuit} &  \textbf{Transpiled circuit} \\
		& & Number of qubits & 4  & 4 \\ 
		H$_2$ & pp-t-UPS(1)/AS(2,2) & Entanglers  & 18  &  29 \\ 
		& & Depth & 28 &  98\\
		\hline
        & & Number of qubits & 8 &  8 \\ 
		H$_2$O & pp-t-UPS(2)/AS(4,4) & Entanglers & 128  &  257 \\ 
		& & Depth & 131  &  414\\
		\bottomrule
	\end{tabular}
\end{table}

\subsection{Energy, gradient, and Hessian method details} \label{subsection:En_Grad_Hes_methods}

In the following, the main results consist of calculating the potential energy surface (PES) and its gradients by contracting and stretching the interatomic distance. In summary, for each geometry i) the $\Theta$-parameters of the oo-VQE scheme were classically optimized, ii) then, these parameters were used in the (pp-)tUPS circuit, yielding the $\ket{0}$ wavefunction, iii) the necessary density elements were measured, constructing the full 1-RDM and 2-RDM matrices, to finally iv) calculate the energy and analytical gradient.

Calculating the nuclear Hessian in Eq. \eqref{eq:NuclearHessian_complet} requires evaluating both its static and its relaxed contributions. First, the static Hessian of Eq. \eqref{eq:static_Hessian}, besides relying on the correct derivatives and orbital bases, depends on the same ingredients as the nuclear gradient: the 1-RDM and 2-RDM matrices. Thus, just as before, the RDMs elements are measured individually to construct the necessary tensors for computing $E^{(2)}_{\text{static}}$. Second, the relaxed contribution of Eq. \eqref{eq:relaxed_Hessian} requires solving the static response equations. To this aim, the construction of the $\boldsymbol{A}$ and $\boldsymbol{B}$ submatrices is achieved through explicit measurements on the quantum hardware \cite{ziems2024options}. Three main reasons justify this choice: i) the SlowQuant program \cite{SlowQuant} includes a ready-to-use linear response algorithm that has proven effective for multireference wavefunctions \cite{ziems2024options,fuglsbjerg2026orbital}. It is therefore expected to perform reasonably well for the pp-tUPS wavefunction; ii) for the hydrogen molecule all necessary (Pauli) excitation operators have already been measured, so no additional measurements are required to construct the hydrogen $\boldsymbol{A}$ and $\boldsymbol{B}$ linear response matrices, and therefore, the orbital Hessian is directly constructed; and iii) for small active spaces, the number of needed expectation values scales better than other methods such as 2D-parameter shift rule \cite{mitarai2018quantum,schuld2019evaluating,izmaylov2021analytic,wierichs2022general,nakagawa2023analytical}. In principle, also the property gradient $f^{(1)}$ can be measured in the quantum hardware. However, this would introduce more sources of errors, so for this work the property gradient is kept classically constructed.

\section{Results and Discussion}
\label{results}

The results are presented in two subsections: the first one corresponds to the analytical gradient simulation in silico and its QPU measurement on real quantum hardware; the second subsection follows the same simulation verification and real experimentation process but for the analytical Hessian.  The simulations resulting from ideal and real quantum hardware are compared to the PySCF and FD references. The measurement details are also summarized within each subsection. 

\subsection{Gradient}

\subsubsection{H$_2$ quantum emulation}

\begin{figure}[H]
	\centering
	\includegraphics[width=0.7\columnwidth]{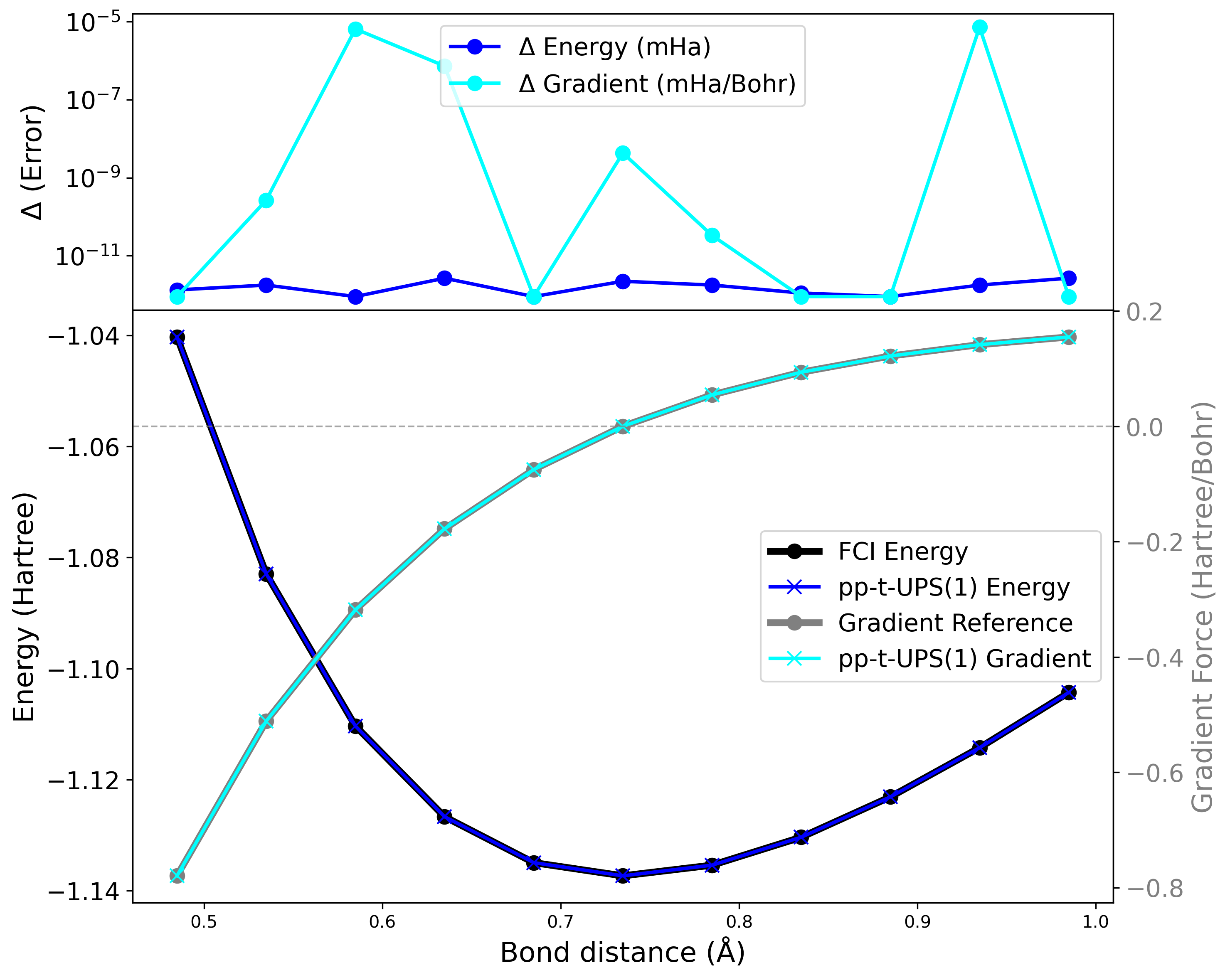}
	\caption{H$_2$ classical FCI and ideal one-layer tUPS energies and gradients. (Bottom) Hydrogen's stretching PES and its associated gradient modulus. (Top) The energy and gradient errors with respect to the classical FCI reference.}
	\label{fig:pptUPS_error}
\end{figure}

After implementing the nuclear gradient equation of Eq. \eqref{eq:explicit_gradiet}, the first step is to verify the performance of ideal simulations. The one-layer tUPS circuit is simulated without gate errors in the infinite shot limit. The PES of the hydrogen molecule is simulated by varying the bond distance, and the results are compared to the FCI PySCF calculations in \figref{fig:pptUPS_error}.
At first glance, the results from the classical FCI reference and the UPS wavefunction are in excellent agreement between each other for the energy and also for the gradient results. The gradient (force) modulus goes through its origin at the equilibrium bond distance as expected. 
Examining in more detail, the differences are small enough, with respect to the desired chemical accuracy of $1.5$ mHa, to consider that the one-layer tUPS and the classical FCI wavefunctions are virtually equivalent. Furthermore, the results coincide with Dalton's gradient calculations up to the $5^\text{th}$ decimal. Thus, in the following results, the FCI results are kept as reference.

\begin{figure}[H]
	\centering
	\includegraphics[width=0.6\columnwidth]{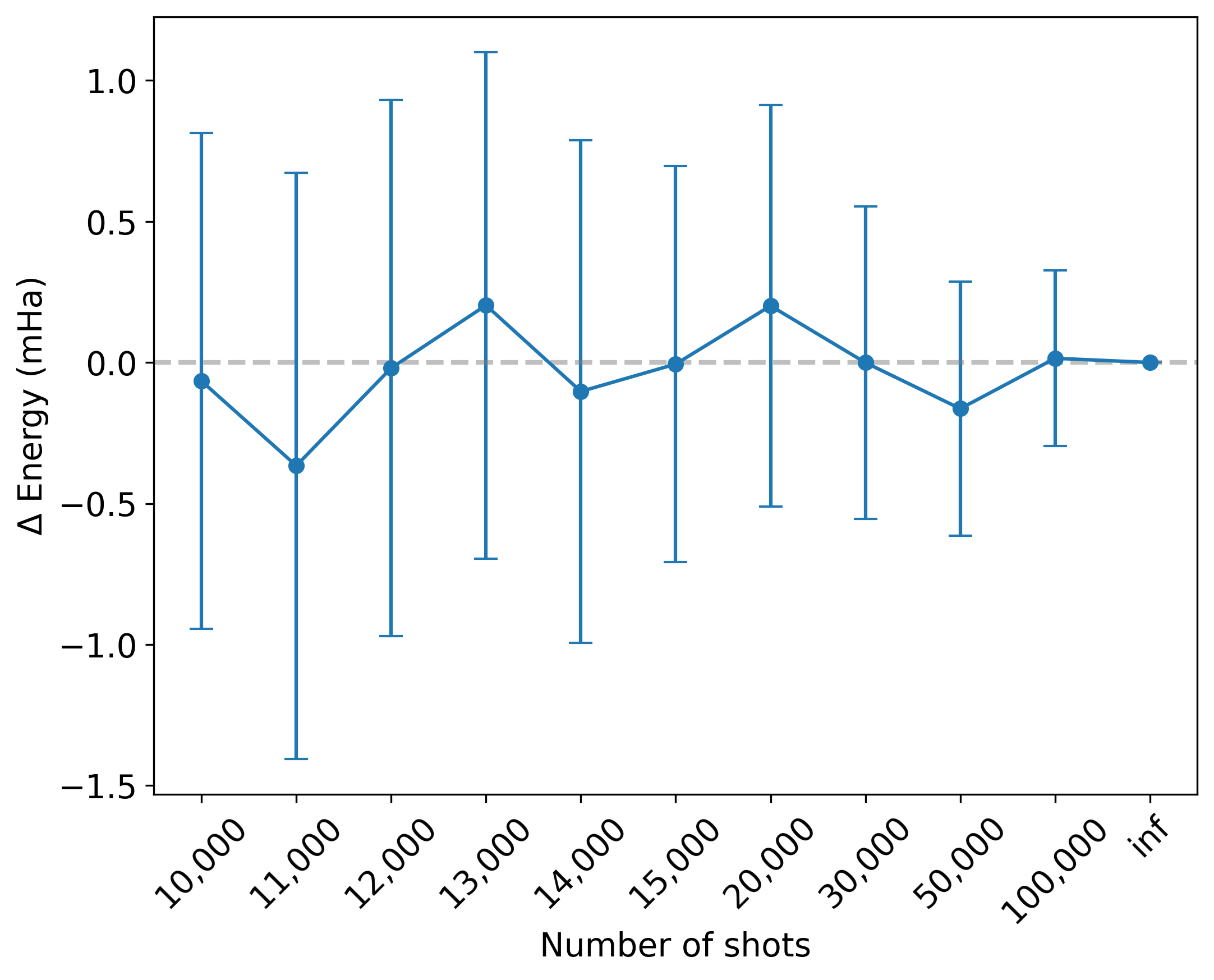}
	\caption{H$_2$/AS(2,2)/STO-3G ground state shot noise simulation. The average and standard deviation of 100 simulations of the electronic energy are shown versus the number of shots. The 'Inf' label stands for an infinite limit of shots. }
	\label{fig:Simulation_onlyShotNoise}
\end{figure}

As a further examination, the effects of shot noise are simulated. Using the equilibrium distance, 100 simulations for different number of shots were performed. The number of shots (10\,000, 11\,000, ..., 100\,000) correspond to the number of measurements per Pauli string. The mean and standard deviations to the ideal energy are plotted in \figref{fig:Simulation_onlyShotNoise}.  
Under the light of these results, the shot noise magnitude is estimated to 1-2 mHa. Hence, shot noise is important only when the results are close to the desired chemical accuracy. 

\subsubsection{H$_2$O quantum emulation}

\begin{figure}[H]
	\centering
	\includegraphics[width=1.0\columnwidth]{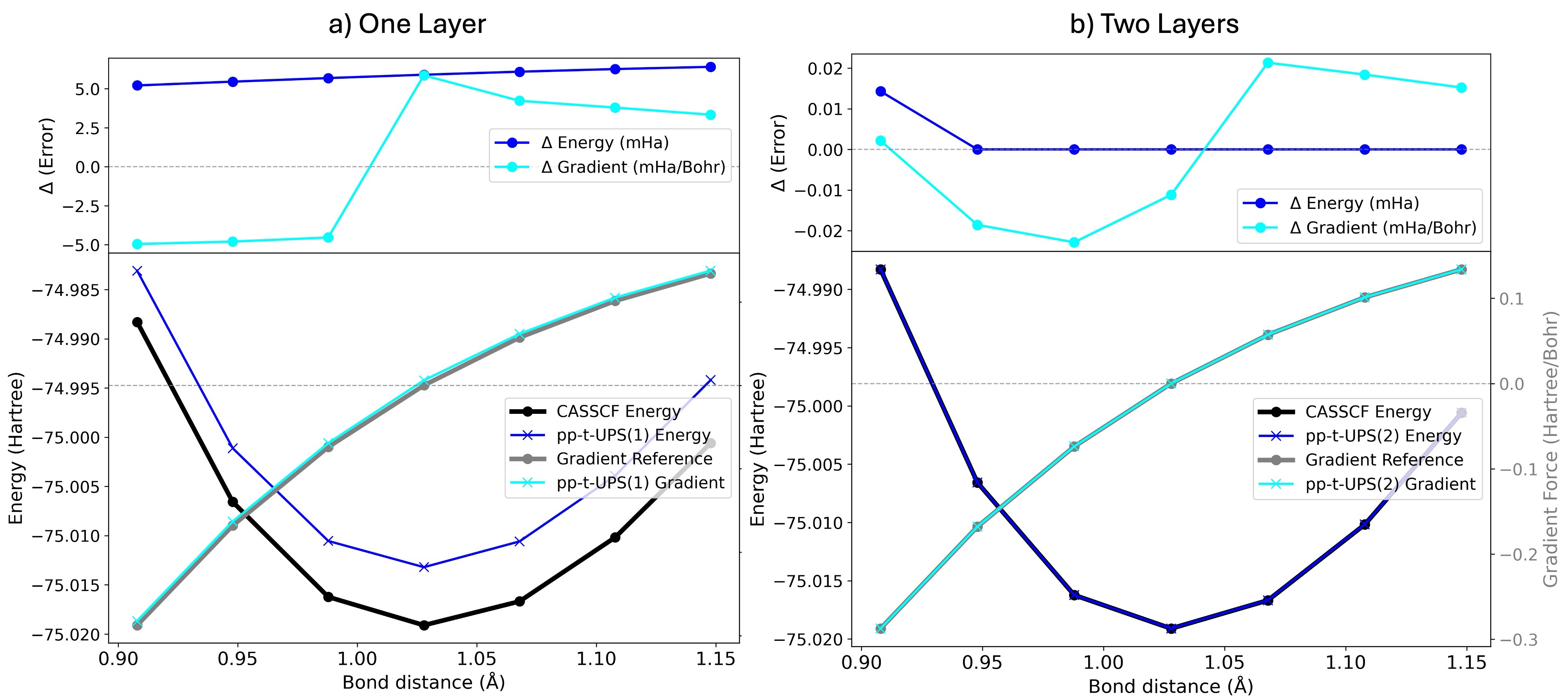}
	\caption{H$_2$O/AS(4,4)/STO-3G classical CASSCF and ideal a) one-layer and b) two-layer tUPS energies and gradients. (Bottom) Water stretching PESs and their associated gradient modulus. (Top) The energy and gradient errors with respect to the classical CASSCF reference. }
	\label{fig:one_two_layer_ideal}
\end{figure}

In the same manner, the ideal (no gate noise and infinite shots) simulation of the water PES is depicted in \figref{fig:one_two_layer_ideal} for one- and two-layers tUPS.
The one-layer tUPS wavefunction does not reproduce the CASSCF values. It has a $5-6$ mHa constant energy overestimation error. Similarly, the simulated gradient modulus underestimates or overestimates the force reference by $5$ mHa/Bohr. On the other hand, the two-layers tUPS wavefunction is virtually identical to the CASSCF wavefunction: energy-wise, only the shortest bond length ($0.9079\text{ \AA}$) differs from the reference by $0.015$ mHa, while the rest of the geometries have a $10^{-7} - 10^{-6}$ mHa error, which is in the order of numerical precision error. Moreover, the gradient forces are always under $0.02$ mHa/Bohr away from the reference. The two-layer tUPS wavefunction is thus considered equivalent to the CASSCF reference for the (4,4) active space. 

\begin{figure}[H]
	\centering
	\includegraphics[width=0.6\columnwidth]{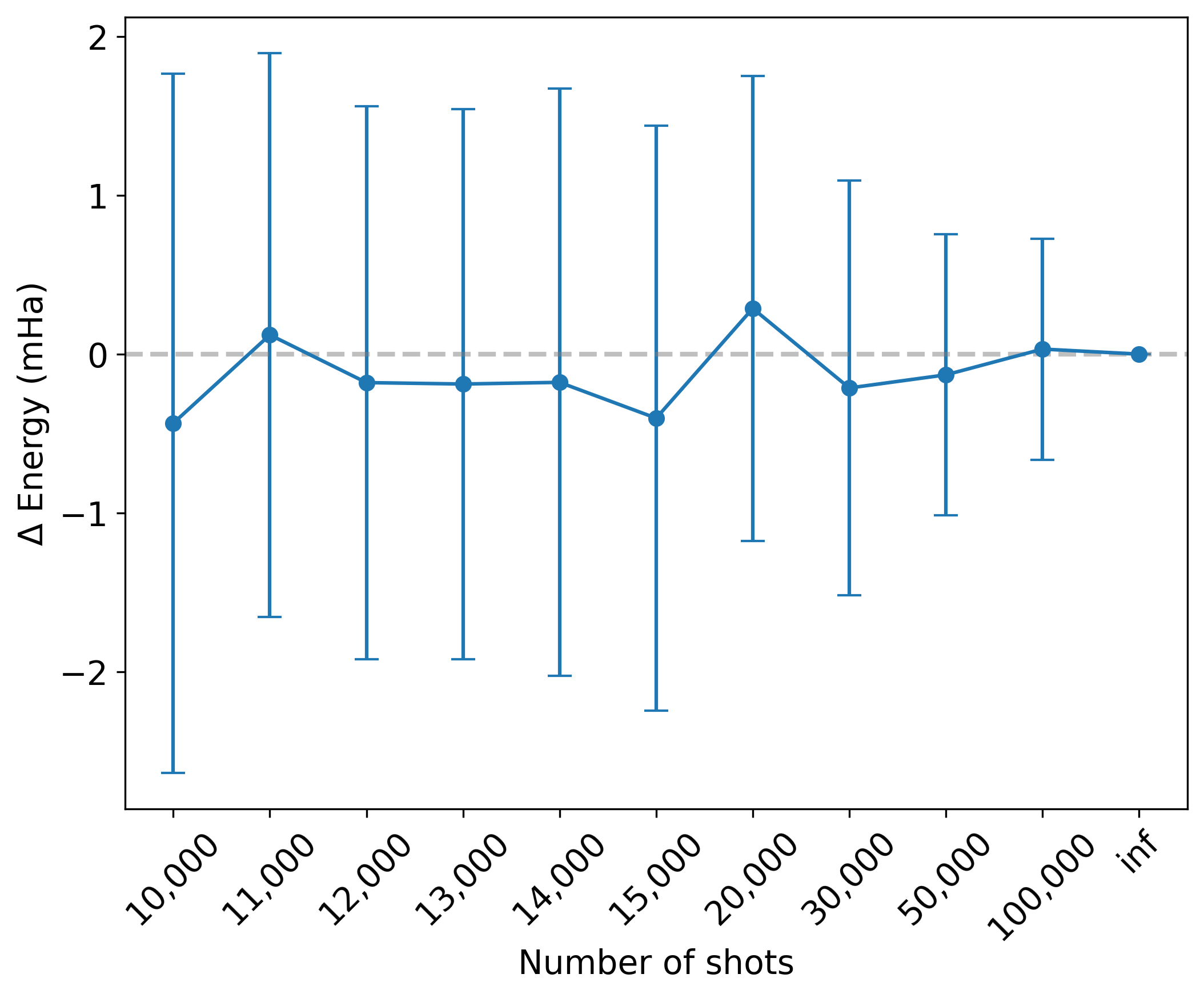}
	\caption{H$_2$O/AS(4,4)/STO-3G ground state shot noise simulation. The average and standard deviation of 100 simulations of the electronic energy are shown versus the number of shots. The 'Inf' label stands for an infinite limit of shots. }
	\label{fig:Simulation_onlyShotNoise_H2O}
\end{figure}

When assessing the shot noise, see \figref{fig:Simulation_onlyShotNoise_H2O}, the error deviation magnitude is around $3-4$ mHa. The standard deviation is almost double with respect to the H$_2$ deviation shot noise. However, the simulation suggests that the mean shot noise error remains under $1$ mHa. Thus, just as before, shot noise error is relevant only on the vicinity of a chemical accuracy precision result. 

\subsubsection{H$_2$ quantum experiment}

Using the same methodology as the in-silico simulation, the calculations are now performed on IBM's Pittsburgh QPU.  Due to the noisy nature of real QPUs, the measured outcome is a noisy raw energy significantly higher than the ideal one (\textit{vide infra}), which necessitates applying the error mitigations outlined above (see subsection \ref{subsection:ErrorMitigation}). We note that despite the frequent hardware calibrations and meticulous error controls, the errors themselves are not stable and can fluctuate over time \cite{rasmussen2025cost}. Therefore, for the repetition of the same calculation, different outcomes and statistical deviations should be expected. 

\begin{figure}[H]
	\centering
	\includegraphics[width=0.7\columnwidth]{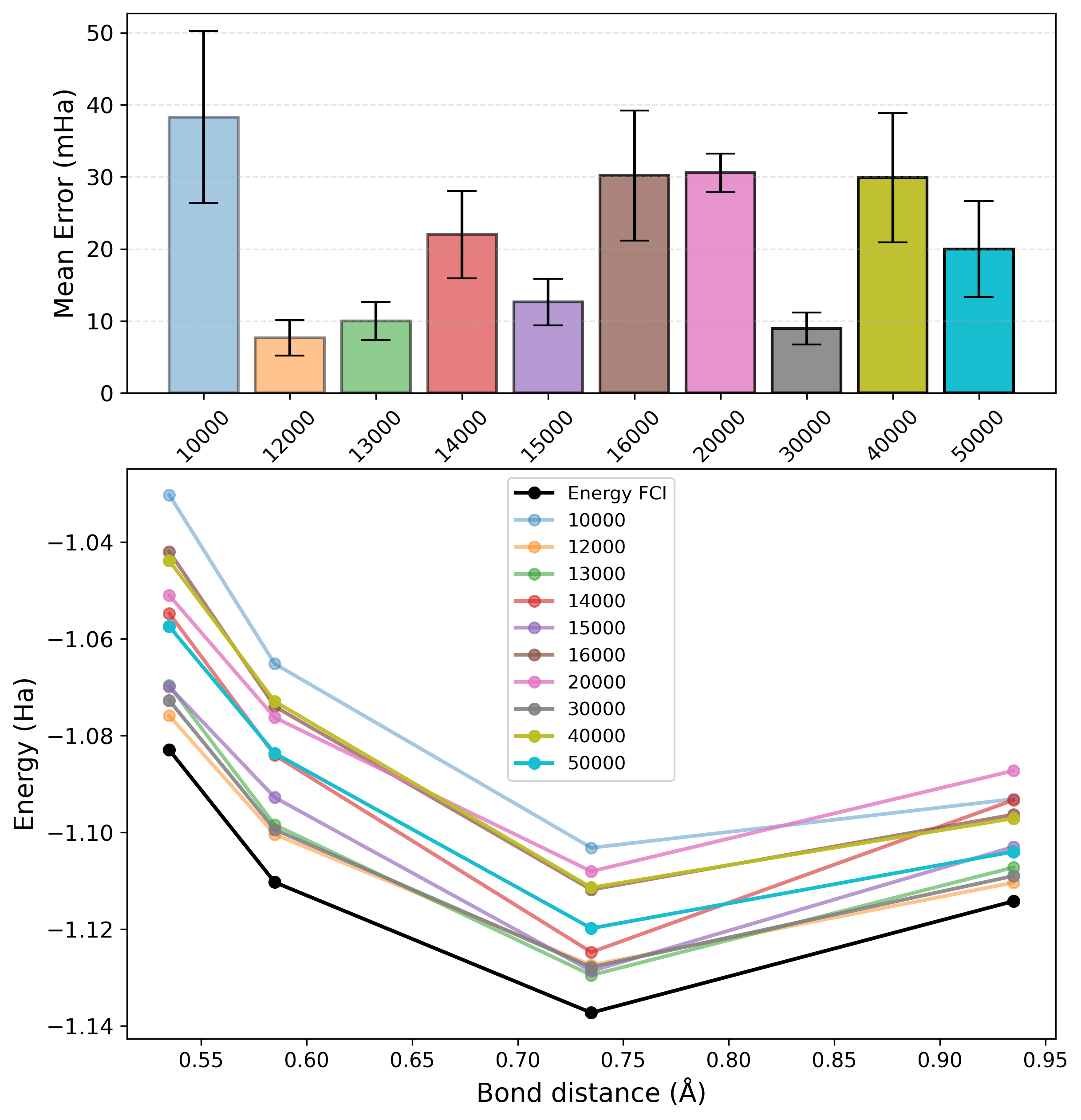}
	\caption{H$_2$/AS(2,2)/STO-3G stretching energies: simulated reference and real measurements on IBM's Pittsburgh QPU using different number of shots. (Bottom) Simulated FCI reference energy (black). The colored lines represent the QPU measurements using different number of shots. (Top) Mean difference and standard deviation between the QPU measurement and the reference for different number of shots.}
	\label{fig:H2_4points_different_shots}
\end{figure}

As a first control, we investigate how the number of requested shots impact the QPU results. Increasing the number of shots refers to increasing the shot budget for each clique Pauli string for the gradient property as well as for each bitstring in the confusion matrix construction.
Only four points in the PES are chosen as representatives for the shot behavior. The mitigated results are depicted in \figref{fig:H2_4points_different_shots}.
Interestingly, the conclusion is that more number of shots does not necessarily correlate to better results. For example, when using 50\,000 shots (cyan), the error is greater for the four geometries, than the 30000 shots (gray). 
%
This can be explained as a QEM inherent competition between: enough shots to characterize the noise, and long calculation time that average over noise drift and fluctuation. For large shot numbers, the QPU time increases and thus the bigger the window for error fluctuations and instabilities. Following these results, the 12000 shot number was chosen for the rest of the calculations. The total necessary shots and QPU time for one geometry are summerized in \tabref{tab:H2_computation_details}.
\begin{table}[H] 
	\centering
	\caption{Hydrogen summary of the QPU measurement for a single geometry. The considered Hamiltonian consists of 41 Pauli groups. \label{tab:H2_computation_details}}
	\begin{tabular}{cccc}
		\toprule
        \multirow{3}{*}{\makecell{Requested \\ number of shots}} & 
		\multirow{3}{*}{\makecell{Total Shots \\ per geometry}} & 
		\multicolumn{2}{c}{\makecell{QPU time ($\sim$min)}} \\
		\cmidrule{3-4}
		& & \makecell{For energy \\ evaluation} & \makecell{For \\ mitigation} \\
		\midrule
        10\,000  & \phantom{0}410 000 &  1 &  1 \\
		12\,000  & \phantom{0}492 000 &  2 &  1 \\
        13\,000  & \phantom{0}533 000 &  2 &  1 \\
        14\,000  & \phantom{0}574 000 &  2 &  1 \\
        15\,000  & \phantom{0}615 000 &  2 &  1 \\
        16\,000  & \phantom{0}656 000 &  2 &  1 \\
        20\,000  & \phantom{0}820 000 &  2 &  2 \\
        30\,000  & 1 230 000 &  3 &  2 \\
        40\,000  & 1 640 000 &  4 &  3 \\
        50\,000  & 2 050 000 &  6 &  4 \\
		\bottomrule
	\end{tabular}
\end{table}

\begin{figure}[H]
	\centering
	\includegraphics[width=0.7\columnwidth]{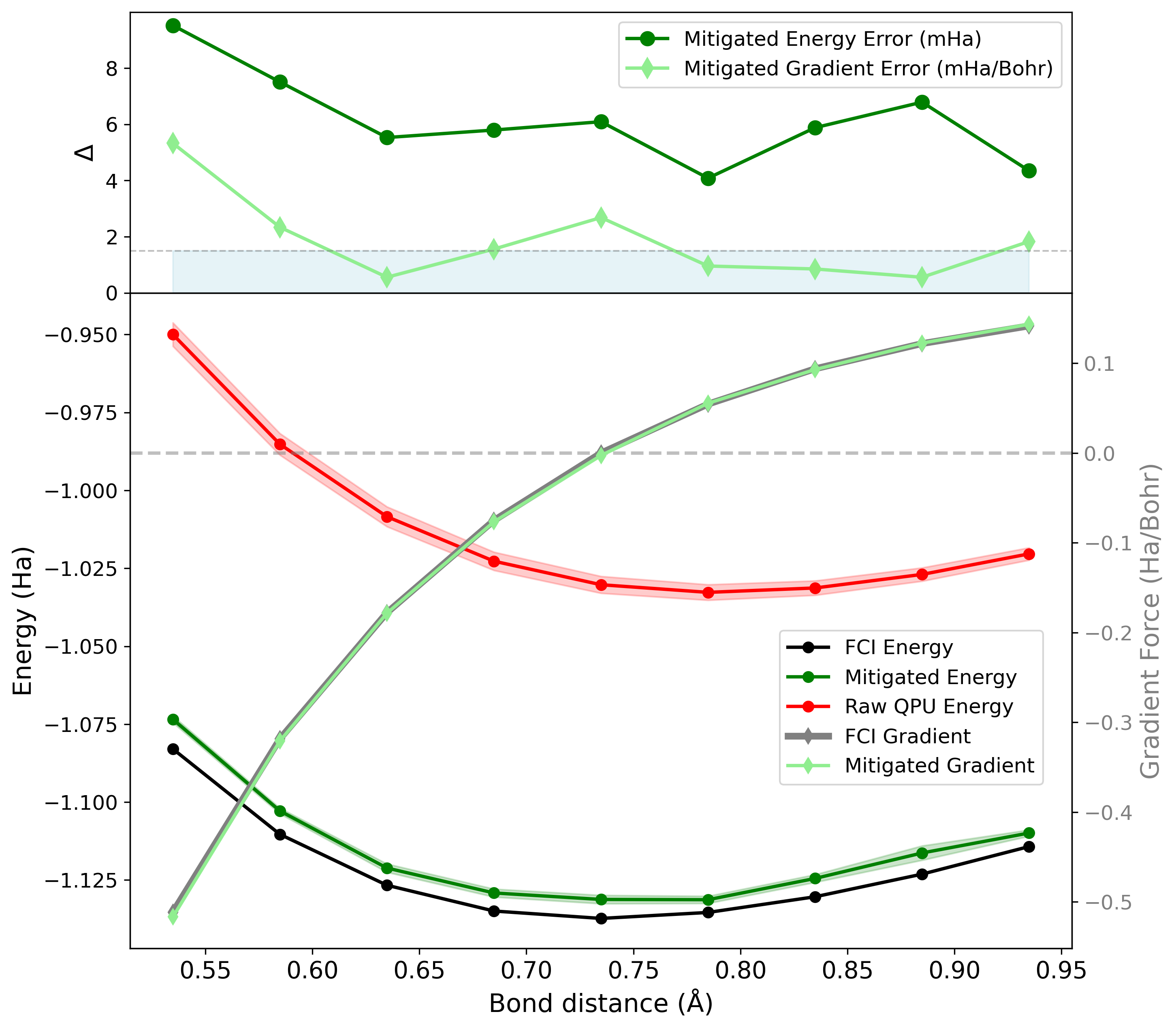}
	\caption{H$_2$/AS(2,2)/STO-3G stretching energies and gradients: simulated reference versus real measurement on IBM's Pittsburgh QPU using 12\,000 shots. (Bottom) Simulated FCI reference energy (black) and gradient (gray). Average raw QPU-measured energy (red). Average mitigated energy (dark green) and gradient (light green) using the M0 and Postselection methods. The standard deviations are shown as shaded red, green, and light green regions. (Top) Difference between the mitigated averaged QPU energy (dark green) and gradient (light green) and their respective reference. The blue shaded area delimits the chemical accuracy region. }
	\label{fig:H2_12000_shots}
\end{figure}

Using the 12\,000 shots scheme, the complete stretching PES is constructed in \figref{fig:H2_12000_shots} using 9 different bond lengths.  For each geometry, 6 identical calculations were performed, retrieving six raw and six mitigated energies, and gradients. The points represent the mean values, while the colored regions delimit their standard deviation. 
The raw results (the QPU energies without mitigation, red) are more than $100$ mHa far from the reference throughout the entire PES. It is two orders of magnitude away from any reliable precision, and the stationary bond distance is shifted by $\sim 1 \Ang$. After mitigation (dark green), the mitigated energies are considerably closer to the FCI reference. The errors after mitigation are below $9$ mHa, remarkably close to chemical accuracy (blue zone) despite the relatively low number of shots. For both, the mitigated energy and gradient curves, the overall shapes are closer to their respective reference. 
The quality of the results is directly related to the quality of the measured density matrices. This error analysis is dealt in subsection \ref{subsubsection:H2_quantum_exp}.
The mitigated gradient errors are below $6$ mHa/Bohr, and some points are even within the chemical accuracy zone. In classical software, such as Gaussian16 \cite{g16}, a 'tight' geometry optimization requires a force threshold of $1.5 \times 10^{-2}$ mHa/Bohr, whereas a standard optimization has a maximum force allowance of $0.45$ mHa/Bohr.  So, the  $\sim 1$ mHa/Bohr precision for the gradient is one order of magnitude higher than a standard classical algorithm, but remain appreciably low.

\subsubsection{H$_2$O quantum experiment}

From the very beginning, the task of simulating the water molecule in quantum hardware announces challenging.
The required QPU resources for the H$_2$O molecule are significantly higher than for the hydrogen molecule, see \tabref{tab:pptUPS_ansatz_details}. For the AS(4,4), 9 times more entanglers, and 4 times the H$_2$ circuit depth are roughly needed. The water Hamiltonian consists of 290 Pauli groups, which is 7 times more than its hydrogen analogous.  
The noise effects are thus expected to be substantially amplified, hence demanding considerably more measurements and time to characterize the noise and the expectation values. As first attempts, just the ground state equilibrium geometry of the water molecule was simulated on the quantum hardware. The three results are summarized in \tabref{tab:H2O_computation_details}.
\begin{table}[H] 
	\centering
	\caption{Water summary of the QPU measurement for the equilibrium geometry. The considered Hamiltonian consists of 290 Pauli groups. \label{tab:H2O_computation_details}}
	\begin{tabular}{ccccccc}
		\toprule
        \multirow{3}{*}{\makecell{Requested \\ number of shots}} & 
		\multirow{3}{*}{\makecell{$\Delta$ Raw Energy \\ (mHa)}} & 
		\multirow{3}{*}{\makecell{$\Delta$ Mitigated Energy \\ (mHa)}} & 
		\multirow{3}{*}{\makecell{Total Shots \\ per geometry}} & 
		\multicolumn{2}{c}{\makecell{QPU time ($\sim$ min)}} \\
		\cmidrule{5-6}
		& & & & \makecell{For energy \\ evaluation} & \makecell{For \\ mitigation} \\
		\midrule
		10 000 & 856 & -201  & 2 900 000 & 14 & 21 \\
		12 000 & 780 & -123 & 3 480 000 & 25 & 32 \\
		30 000 & 880 & -164 & 8 700 000 & 120 & 125 \\
		\bottomrule
	\end{tabular}
\end{table}
The presented results are single evaluations (conversely to the 5 batches of runs for the hydrogen), so their performance related to the retrieved energy error should be taken as tendencies instead of absolute values. The first takeaway is that the QPU time quickly upsurges with the number of shots per Pauli group. The QPU resources are limited, therefore, such long calculations are already an important bottleneck for the method. The second remark is that the energy error does not necessarily decrease with the number of shots. Reminiscent to the hydrogen shot study, this is probably related to the backfires of long evaluation time: increasing the number of shots, entails a high price on QPU time, opening the possibilities to error drift and noise mischaracterization. In this case, the error profile is overestimated, meaning the mitigated energies underestimate the aimed values by about 150 mHa. Nevertheless, this still represents a 4- to 6-fold improvement over the raw energies.
The overall errors are expected to lessen with the continuous improvement of qubit and gate performances. However, the computer resources for error characterization still exponentially grows with the size of the system, as the number of possible determinants quadratically grows with the number of qubits. For a more detailed explanation the reader is referred to ref.~\cite{rasmussen2025cost}.
These results are clear examples of the current challenges of long circuit simulations and the scaling limitations of the mitigation method. 

\subsection{Hessian}

The remainder of calculations in this subsection are performed by considering the hydrogen molecule at the equilibrium geometry. Given that the gradient calculations for the water molecule are already too costly, this molecule is no longer studied.


\subsubsection{H$_2$ Simulation and verification}

As described in subsection \ref{subsection:En_Grad_Hes_methods}, the PySCF reference is built using FD. This classical matrix serves as the Hessian reference, and its diagonalization result in the reference vibrational frequencies.
Its explicit values can be seen in SI. 
Secondly, the ideal tUPS wavefunction is used to evaluate the ideal tUPS nuclear Hessian: the 1-RDM, the 2-RDM, and the electronic Hessian are constructed using the simulated tUPS wavefunction, and the analytical expression is applied. The ideal tUPS nuclear Hessian is also explicitly written in SI. The ideal PySCF and UPS Hessians are close, differing at most by $\sim 1$ mHa/Bohr$^{2}$ in some elements. Let us first compare the resulting vibrational frequencies of each nuclear Hessian. The two nuclear Hessians are transformed into their mass-weighted form, and are afterwards diagonalized. Among the six eigenvalues, the last one correspond to the stretching normal mode. It is found that the FD and the ideal tUPS associated vibrational frequencies are $5000~$cm$^\text{-1}$ and $4993~$cm$^\text{-1}$, respectively. A $7$ cm$^\text{-1}$ frequency difference is appropriately small, an error of about $0.15 \%$, given the level of approximation between methods. Thence, it is considered safe to proceed using this methodology, as it is expected that the vibrational frequency discrepancies will be considerably larger when using the imperfect real quantum hardware.

~ 




\subsubsection{H$_2$ quantum Experiment} \label{subsubsection:H2_quantum_exp}

\begin{figure}[H]
	\centering
	\includegraphics[width=0.9\columnwidth]{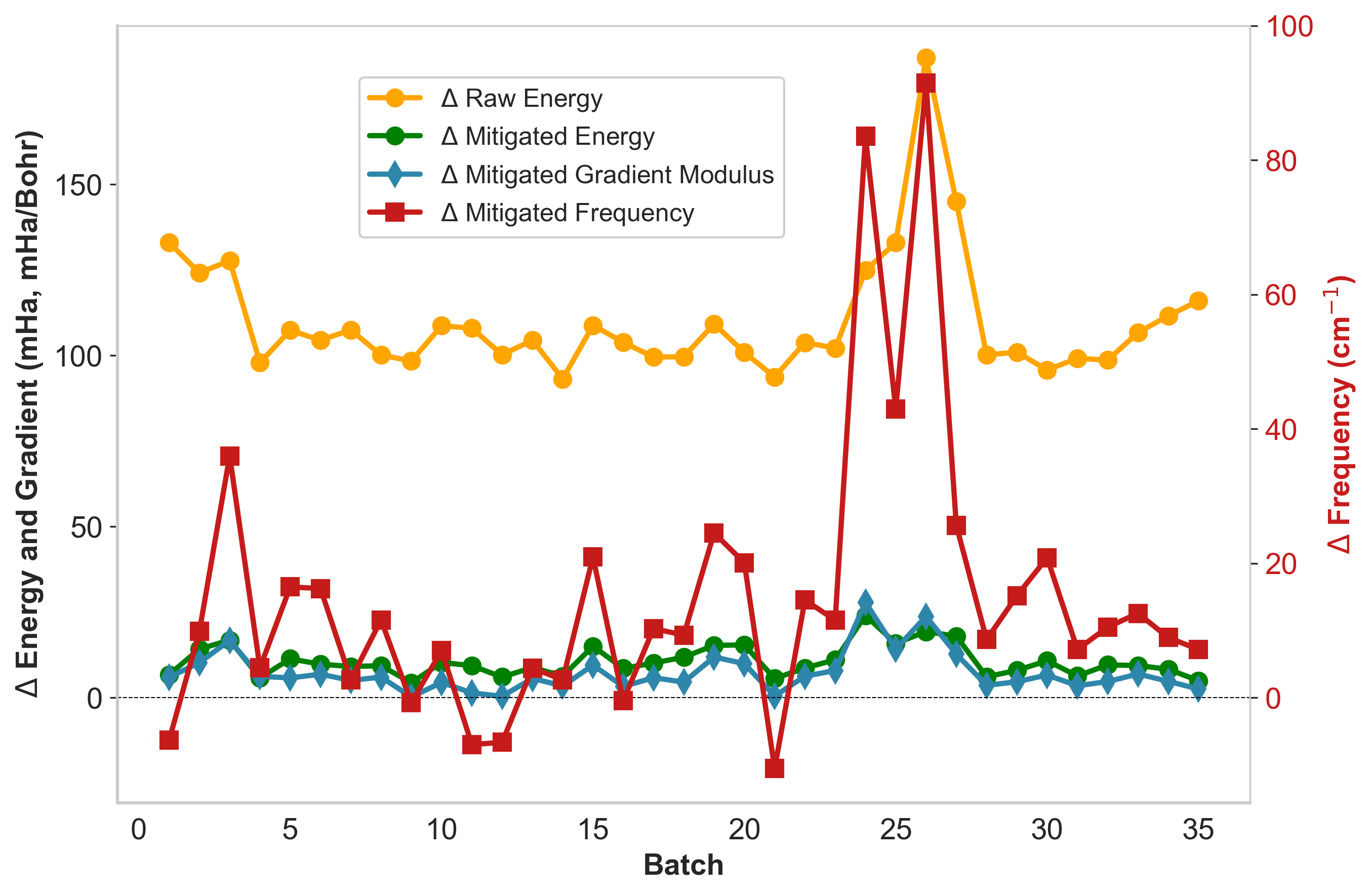}
	\caption{The 35 measurement batches of the ground state geometry properties performed on IBM's Pittsburgh QPU backend. The energies, gradient, and frequencies are compared to the ideal fermionic wavefunction results.}
	\label{fig:Hessian_12000_shots}
\end{figure}

In the same manner as before, the optimized tUPS ansatz is applied in the quantum hardware. The resulting QPU wavefunction is used to measure the RDMs and electronic Hessian. The analytical Hessian is then calculated, and upon diagonalization, the eigen frequencies are retrieved. This exact procedure for determining the nuclear Hessian is repeated 46 times, from which only 35 batches are retained. The dropped batches were run on unreliable qubits with high error rates and are therefore considered deficient. 
The read out calibration of the qubits used is shown in SI. The results of the raw energies, the mitigated energies, and the associated gradient moduli and vibrational frequencies are depicted in \figref{fig:Hessian_12000_shots}.
Similar to the gradient results, the raw energies are on average $110$ mHa far from the reference. After mitigation, the energies overestimate the reference by $11$ mHa on average. Even for batches with exceptionally big errors (\textit{e.g.} batches 25, 26, and 27), the $\bm M_0$ error mitigation is capable of characterizing and correcting the error lowering the discrepancy below $24$ mHa. From the mitigated results, the retrieved gradient moduli are on average $7$ mHa/Bohr above the ideal module-less gradient. The vibrational frequencies are more sensitive to error and thus susceptible to variations. To better visualize this, the 6 resulting frequencies are summarized in \tabref{tab:H2_VibrationalFreq_results}.
\begin{table}[H] 
	\centering
	\caption{Eigen frequencies (cm$^{-1}$) of the classical FD reference, the ideal analytical tUPS, and the measured Hessians.  \label{tab:H2_VibrationalFreq_results}}
	\begin{tabular}{ccccc}
		\toprule
		Eigenvalue & \makecell{Ideal tUPS \\ wavefunction} & \makecell{QPU tUPS \\ wavefunction} & \makecell{Finite difference \\ reference} & Type \\ 
		\hline
		  1 & 0 & -407 $\pm$ 171 & 0 & x- Translation \\
		 2 & 0 & -407 $\pm$ 171 & 0 & y- Translation  \\
		 3 & 0 & \hspace{0.12cm} 0 $\pm$ 0 & 0 & z- Translation  \\
		 4 & 34 & \hspace{0.23cm} 2 $\pm$ 14 & 35 & x- Rotation \\
		 5 & 34 & \hspace{0.25cm} 2 $\pm$ 14 & 35 & y- Rotation \\
		 6 & 4993 &  \hspace{-0.33cm}  5008 $\pm$ 21 & 5000 & Normal mode stretching \\
		\bottomrule
	\end{tabular}
\end{table}
The FD reference is, once again, the numerical PySCF reference which was further validated by a Dalton calculation; and the ideal tUPS is the aimed analytical result. The last eigenvalue corresponds to the stretching vibrational frequency. The retrieved experimental value is $5008$ cm$^{-1}$, which is considerably close to the aimed value. Frequency errors in the order of $10$ cm$^{-1}$ are not particularly significant, as frequencies are sensitive to instabilities, or even just basis-set choices. For instance, as a control, the PySCF FD was repeated using the 3-21G Pople basis, giving a vibrational frequency of $4595$ cm$^{-1}$, illustrating thus the relative sensibility of these values. However, remarkable discrepancies are found for the other eigenvalues. Despite that these values are not the researched properties, these are witness of the overall quality of the retrieved Hessians. Specially, the first two eigenvalues associated to pure translations show negative values, and thus instabilities within the matrix. On \tabref{tab:properties_comparison} the tensor errors are summarized and their explicit writing is in SI.
\begin{table}[h]
	\centering
	\caption{Property errors with respect to the ideal tUPS reference.}
	\begin{tabular}{l r@{$\,\pm\,$}ll}
		\hline
		\textbf{Property} & \multicolumn{2}{c}{\textbf{Mean $\pm$ Std}} & \textbf{Unit} \\
		\hline
		$\Delta$ Ground State Energy                  & $10.593$                  & $4.469$ & mHa \\
		$\Delta$ Gradient Modulus     & $7.261$                  & $5.894$ & mHa/Bohr \\
		$\Delta$ 1-RDM Max         & $15.571$                 & $6.543$ \\
		$\Delta$ 1-RDM Distance        & $23.901$                 & $9.350$ \\
		$\Delta$ 2-RDM Max         & $22.016$                 & $5.820$ \\
		$\Delta$ 2-RDM Distance        & $46.238$                 & $10.642$ \\
		$\Delta A$ Max       & $18.524$                  & $7.911$  & mHa\\
		$\Delta A$ Distance      & $28.829$                  & $9.073$  & mHa\\
		$\Delta B$ Max       & $4.646$                  & $1.470$  & mHa\\
		$\Delta B$ Distance      & $6.895$                  & $2.062$  & mHa\\
		$\Delta \G^{(0)}$ Max      & $19.938$                  & $7.980$  & mHa\\
		$\Delta \G^{(0)}$  Distance    & $29.492$                  & $9.245$  & mHa\\
		$\Delta$ Nuclear Hessian Max         & $4.106$                  & $3.632$ & mHa/Bohr$^2$\\
		$\Delta$ Nuclear Hessian Distance       & $12.641$                  & $11.081$  & mHa/Bohr$^2$\\
		\hline
	\end{tabular}
	\label{tab:properties_comparison}
\end{table}
The different matrix references are subtracted to the retrieved matrices, and then two type of errors are depicted: the Max which is naturally the largest matrix element, and the distance which follows Eq.~\ref{eq:Grad_force}, giving a value of the overall discrepancy. It is difficult to pinpoint a single culprit for the instabilities or errors; nevertheless, some tendencies can be identified: there is no linear relation between the electronic Hessian $\G^{(0)}$ and the linear response matrices that compose it; the largest source of error appears to affect the 2-RDM, yet the total Hessian error remains considerably lower. This presumably points to error cancellations and a noticeably intertwined behavior among the contributing terms.  
This error magnitude is comparable to other modern results \cite{ziems2024options,rasmussen2025cost}, for instance, ref.~\cite{reinholdt2025critical} also finds energy error on the order of $10$ mHa for small systems, using up to $10^6$ shots.
Improving these results require more adapted and accurate error techniques \cite{rasmussen2025cost, aharonov2025reliable}, or the much expected fault-tolerant hardware.

\section{Conclusions and Outlook}
\label{outlook}

The PES and analytical nuclear gradient were calculated for the hydrogen and water molecules using state-vector simulation and real quantum hardware emulation. The method was based on measuring the individual components of the one-particle and two-particle density matrices, using the optimized (pp-)tUPS ansatz embedded within the oo-VQE frame and different active spaces. The $\bm M_0$ error mitigation and Post-Selection strategies were applied to correct the QPU raw measurements, considerably improving the energies and geometrical gradient. While the hydrogen molecule results required only few minutes using the real quantum hardware to fairly approach the ideal references, the water molecule needed substantially more computational resources and still underestimating the energies.

The study continued focusing on the H$_2$ Hessian simulation and real experimentation. Classical numerical FD and ideal analytical implementations were used as reference, concluding that they were virtually equivalent for the chosen ansatz. Then, the quantum emulation was repeated 35 times for statistical significance. The retrieved vibrational frequency associated to the stretching normal mode was found adequately close to the references. The other Hessian eigenvalues and the measured matrices were discussed, quantified and analyzed. The conclusion pointing towards big measurement errors within the 2-RDM and the $\boldsymbol{A}$ linear response matrix necessary to compute the electronic Hessian. 

It most be noted that the linear response framework was used to calculate the electronic Hessian, which is particularly useful for time-dependent properties. Nonetheless, vibrational frequencies are static properties that might benefit of different (quantum) methods such as the parameter shift rule \cite{mitarai2018quantum,schuld2019evaluating,izmaylov2021analytic,wierichs2022general,nakagawa2023analytical}, that uses the analytical expression of the required orbital second derivatives. However, for the small case of H$_2$, using the parameter shift rule would require more QPU evaluations, whereas the linear response algorithm needs no extra measurements as all the necessary operators where already measured during the energy measurements. Applying the 2D-parameter shift rule to bigger systems and properties is to be delegated for future work.  

Finally, the work proposed herein is already an example of the existing applications and capacities of current quantum hardware for chemistry applications. This same methodology is suitable for near-term quantum computers as it yields compact equations and controlled measurements for small systems. Future works will focus on retrieving infrared spectra within active spaces, implementing the Hessian-vector multiplication for the Davidson resolution \cite{davidson_1975} to avoid explicit measurement of the full electronic Hessian, and better mitigation techniques for the used tiled ansatz.

\section{Acknowledgments}

The authors are grateful for fruitful discussions with Hans Jørgen Aa. Jensen. Financial support from the Novo Nordisk Foundation (NNF) for the focused research project ``Hybrid Quantum Chemistry on Hybrid Quantum Computers'' (NNF Grant No. NNFSA220080996) and from Innovation Fund Denmark for the Eureka Project ``Q-Chemion'' (4340-00006B) is acknowledged. K.M.Z. acknowledges financial support from the Royal Society of Chemistry Collaboration grant, C25-1492721325.

\clearpage
\section{Supporting Information}

\renewcommand*{\thefigure}{S\arabic{figure}}
\setcounter{figure}{0}  
\setcounter{table}{0}

\subsection{Mathematical Formulae}

The nuclear-nuclear terms:
\begin{equation}
    E_\text{NN}(\geo) = \sum_{A\neq B} \frac{Z_A Z_B}{\abs{\geo_A - \geo_B}}
\end{equation}
\begin{equation}
  E^{(1)}_{\text{NN}, ~A\alpha} =  \frac{\partial E_\text{NN}}{\partial R_{A\alpha}} = -\sum_{A \neq B} Z_A Z_B \frac{R_{A\alpha} - R_{B\alpha}}{|\mathbf{R}_A - \mathbf{R}_B|^3}
\end{equation}
\begin{equation}
E^{(2)}_{\text{NN}, ~A\alpha B\beta} =  \frac{\partial^2 E_\text{NN}}{\partial R_{A\alpha} \, \partial R_{B\beta}} = 
\begin{cases}
Z_A Z_B \left( \frac{\delta_{\alpha\beta}}{|\mathbf{R}_{AB}|^3} - \frac{3 \, (R_{A\alpha} - R_{B\alpha})(R_{A\beta} - R_{B\beta})}{|\mathbf{R}_{AB}|^5} \right) & A \neq B \\[10pt]
-\sum_{A \neq B'} Z_A Z_{B'} \left( \frac{\delta_{\alpha\beta}}{|\mathbf{R}_{AB'}|^3} - \frac{3 \, (R_{A\alpha} - R_{B'\alpha})(R_{A\beta} - R_{B'\beta})}{|\mathbf{R}_{AB'}|^5} \right) & A = B
\end{cases}
\end{equation}
The one- and two-particle reduced density matrices in the AO basis:
\begin{equation}
 D_{\text{ao}~\mu\nu}  = \bra{\text{MC}} \Es_{\mu\nu} \ket{\text{MC}} = \bra{\text{MC}} \sum_\sigma \adag_{\mu\sigma} \aan_{\nu\sigma} \ket{\text{MC}}
\end{equation}
\begin{equation}
 d_{\text{ao}~\mu\nu \lambda \sigma}  = \bra{\text{MC}} \Ed_{\mu\nu \lambda \sigma} \ket{\text{MC}} = \bra{\text{MC}}  \Es_{\mu\nu} \Es_{\lambda \sigma} - \delta_{\lambda \nu}\Es_{\mu \sigma}\ket{\text{MC}}
\end{equation}
The effective density matrices:
\begin{equation}
	F^{[0]}_{\text{ao}~ A\alpha, \mu\nu} = \sum_b D_{\mu b} h^{(0)}_{\text{AO}~A\alpha, \nu b} + \sum_{b g d} d_{\mu bgd}  g^{(0)}_{\text{AO}~ A\alpha, \nu bgd}
\end{equation}
\begin{equation}
	F^{[1]}_{A\alpha,pq} = \sum_b D_{pb} h^{(1)}_{A\alpha,qb} + \sum_{b g d} d_{pbgd}  g^{(1)}_{A\alpha, qbgd}
\end{equation}
The first derivatives of the electron integrals and the overlap in the AO basis:
\begin{equation}
		\begin{aligned}
			h^{(1)}_{\text{AO}~ A\alpha, \mu\nu} = &\, \delta_{\mu \in A} \bra{\frac{\partial \chi_\mu}{\partial R_{A\alpha}}} h \ket{\chi_\nu} + \delta_{\nu \in A} \bra{\chi_\mu} h \ket{\frac{\partial \chi_\nu}{\partial R_{A\alpha}}} \\
			&+ \bra{\chi_\mu} -Z_A \frac{\partial}{\partial R_{A\alpha}} \frac{1}{|\mathbf{r} - \mathbf{R}_A|} \ket{\chi_\nu}
		\end{aligned}
\end{equation}
\vspace*{0.5cm}
\begin{equation}
		\begin{aligned}
			g_{\text{AO}~ A\alpha, \mu\nu\lambda\sigma}^{(1)}
			=
			\frac{\partial g_{\mu\nu\lambda\sigma}}{\partial R_{A\alpha}} = &\, \delta_{\mu \in A} \bra{\frac{\partial \chi_\mu(1)}{\partial R_{A\alpha}}\chi_\nu(2)} \frac{1}{r_{12}} \ket{\chi_\lambda(1)\chi_\sigma(2)} \\
			&+ \delta_{\nu \in A} \bra{\chi_\mu(1)\frac{\partial \chi_\nu(2)}{\partial R_{A\alpha}}} \frac{1}{r_{12}} \ket{\chi_\lambda(1)\chi_\sigma(2)} \\
			&+ \delta_{\lambda \in A} \bra{\chi_\mu(1)\chi_\nu(2)} \frac{1}{r_{12}} \ket{\frac{\partial \chi_\lambda(1)}{\partial R_{A\alpha}}\chi_\sigma(2)} \\
			&+ \delta_{\sigma \in A} \bra{\chi_\mu(1)\chi_\nu(2)} \frac{1}{r_{12}} \ket{\chi_\lambda(1)\frac{\partial \chi_\sigma(2)}{\partial R_{A\alpha}}}
		\end{aligned}
\end{equation}
\begin{equation}
		S_{\text{AO}~A\alpha, \mu\nu}^{(1)} =	\frac{\partial}{\partial R_{A\alpha}}  \braket{\chi_\mu}{\chi_\nu} =   \delta_{\mu \in A}\braket{ 	\frac{\partial \chi_\mu}{\partial R_{A\alpha}}  }{\chi_\nu} +  \delta_{\nu \in A} \braket{\chi_\mu}{	\frac{\partial \chi_\nu}{\partial R_{A\alpha}} } 
\end{equation}
The metric correction in the MO basis:
\begin{equation}
	\begin{aligned}
		(S^{(1)}, F^{[0]})_{B\beta, pq} &= \sum_b D_{pb} \{S^{(1)},h^{(0)}\}_{B\beta, qb} + \sum_{bgd} d_{pbgd}  \{S^{(1)}, g^{(0)}\}_{B\beta,qbgd}  \\
		\{S^{(1)},h^{(0)}\}_{B\beta, pq} &= \sum_a (S^{(1)}_{B\beta, pa} h^{(0)}_{a q} + S^{(1)}_{B\beta, qa} h^{(0)}_{pa}) \\
		\{S^{(1)},g^{(0)}\}_{B\beta, pqrs} &= \sum_a  (S^{(1)}_{B\beta, pa} g^{(0)}_{a qrs} + S^{(1)}_{B\beta, qa} g^{(0)}_{pa rs}+ S^{(1)}_{B\beta, ra} g^{(0)}_{pqa s} + S^{(1)}_{B\beta, sa} g^{(0)}_{pqra } )
	\end{aligned}
\end{equation}
The second derivatives of the electron integrals and the overlap in the AO basis:
\begin{equation}
		\begin{aligned}
			h^{(2)}_{\text{AO}~A\alpha B\beta, \mu\nu} = &\, \delta_{\mu \in A} \delta_{\mu \in B} \bra{\frac{\partial^2 \chi_\mu}{\partial R_{A\alpha} \partial R_{B\beta}}} h \ket{\chi_\nu} \\
			&+ \delta_{\nu \in A} \delta_{\nu \in B} \bra{\chi_\mu} h \ket{\frac{\partial^2 \chi_\nu}{\partial R_{A\alpha} \partial R_{B\beta}}} \\
			&+ \delta_{\mu \in A} \delta_{\nu \in B} \bra{\frac{\partial \chi_\mu}{\partial R_{A\alpha}}} h \ket{\frac{\partial \chi_\nu}{\partial R_{B\beta}}} \\
			&+ \delta_{\mu \in B} \delta_{\nu \in A} \bra{\frac{\partial \chi_\mu}{\partial R_{B\beta}}} h \ket{\frac{\partial \chi_\nu}{\partial R_{A\alpha}}} \\
			&+ \delta_{\mu \in A} \bra{\frac{\partial \chi_\mu}{\partial R_{A\alpha}}} \left(-Z_B \frac{\partial}{\partial R_{B\beta}} \frac{1}{|\mathbf{r} - \mathbf{R}_B|}\right) \ket{\chi_\nu} \\
			&+ \delta_{\nu \in A} \bra{\chi_\mu} \left(-Z_B \frac{\partial}{\partial R_{B\beta}} \frac{1}{|\mathbf{r} - \mathbf{R}_B|}\right) \ket{\frac{\partial \chi_\nu}{\partial R_{A\alpha}}} \\
			&+ \delta_{\mu \in B} \bra{\frac{\partial \chi_\mu}{\partial R_{B\beta}}} \left(-Z_A \frac{\partial}{\partial R_{A\alpha}} \frac{1}{|\mathbf{r} - \mathbf{R}_A|}\right) \ket{\chi_\nu} \\
			&+ \delta_{\nu \in B} \bra{\chi_\mu} \left(-Z_A \frac{\partial}{\partial R_{A\alpha}} \frac{1}{|\mathbf{r} - \mathbf{R}_A|}\right) \ket{\frac{\partial \chi_\nu}{\partial R_{B\beta}}} \\
			&+ \delta_{A=B} \bra{\chi_\mu} -Z_A \frac{\partial^2}{\partial R_{A\alpha} \partial R_{A\beta}} \frac{1}{|\mathbf{r} - \mathbf{R}_A|} \ket{\chi_\nu}
		\end{aligned}
\end{equation}
\begin{equation}
		\begin{aligned}
			g^{(2)}_{\text{AO}~A\alpha B\beta, \mu\nu\lambda\sigma} =& \frac{\partial^2 g_{\mu\nu\lambda\sigma}}{\partial R_{A\alpha} \partial R_{B\beta}} \\
			=&\hspace{0.4cm} \delta_{\mu \in A} \delta_{\mu \in B} \bra{\frac{\partial^2 \chi_\mu(1)}{\partial R_{A\alpha} \partial R_{B\beta}}\chi_\nu(2)} \frac{1}{r_{12}} \ket{\chi_\lambda(1)\chi_\sigma(2)} \\
			&+ \delta_{\nu \in A} \delta_{\nu \in B} \bra{\chi_\mu(1)\frac{\partial^2 \chi_\nu(2)}{\partial R_{A\alpha} \partial R_{B\beta}}} \frac{1}{r_{12}} \ket{\chi_\lambda(1)\chi_\sigma(2)} \\
			&+ \delta_{\lambda \in A} \delta_{\lambda \in B} \bra{\chi_\mu(1)\chi_\nu(2)} \frac{1}{r_{12}} \ket{\frac{\partial^2 \chi_\lambda(1)}{\partial R_{A\alpha} \partial R_{B\beta}}\chi_\sigma(2)} \\
			&+ \delta_{\sigma \in A} \delta_{\sigma \in B} \bra{\chi_\mu(1)\chi_\nu(2)} \frac{1}{r_{12}} \ket{\chi_\lambda(1)\frac{\partial^2 \chi_\sigma(2)}{\partial R_{A\alpha} \partial R_{B\beta}}} \\
			&+ \delta_{\mu \in A} \delta_{\nu \in B} \left<\frac{\partial \mu}{\partial R_{A\alpha}} \frac{\partial \nu}{\partial R_{B\beta}} | \lambda \sigma \right> \\
			&+ \delta_{\mu \in A} \delta_{\lambda \in B} \left<\frac{\partial \mu}{\partial R_{A\alpha}} \nu | \frac{\partial \lambda}{\partial R_{B\beta}} \sigma \right> \\
			&+ \delta_{\mu \in A} \delta_{\sigma \in B} \left<\frac{\partial \mu}{\partial R_{A\alpha}} \nu | \lambda \frac{\partial \sigma}{\partial R_{B\beta}} \right> \\
			&+ \delta_{\nu \in A} \delta_{\mu \in B} \left<\frac{\partial \mu}{\partial R_{B\beta}} \frac{\partial \nu}{\partial R_{A\alpha}} | \lambda \sigma \right> \\
			&+ \delta_{\nu \in A} \delta_{\lambda \in B} \left<\mu \frac{\partial \nu}{\partial R_{A\alpha}} | \frac{\partial \lambda}{\partial R_{B\beta}} \sigma \right> \\
			&+ \delta_{\nu \in A} \delta_{\sigma \in B} \left<\mu \frac{\partial \nu}{\partial R_{A\alpha}} | \lambda \frac{\partial \sigma}{\partial R_{B\beta}} \right> \\
			&+ \delta_{\lambda \in A} \delta_{\mu \in B} \left<\frac{\partial \mu}{\partial R_{B\beta}} \nu | \frac{\partial \lambda}{\partial R_{A\alpha}} \sigma \right> \\
			&+ \delta_{\lambda \in A} \delta_{\nu \in B} \left<\mu \frac{\partial \nu}{\partial R_{B\beta}} | \frac{\partial \lambda}{\partial R_{A\alpha}} \sigma \right> \\
			&+ \delta_{\lambda \in A} \delta_{\sigma \in B} \left<\mu \nu | \frac{\partial \lambda}{\partial R_{A\alpha}} \frac{\partial \sigma}{\partial R_{B\beta}} \right> \\
			&+ \delta_{\sigma \in A} \delta_{\mu \in B} \left<\frac{\partial \mu}{\partial R_{B\beta}} \nu | \lambda \frac{\partial \sigma}{\partial R_{A\alpha}} \right> \\
			&+ \delta_{\sigma \in A} \delta_{\nu \in B} \left<\mu \frac{\partial \nu}{\partial R_{B\beta}} | \lambda \frac{\partial \sigma}{\partial R_{A\alpha}} \right> \\
			&+ \delta_{\sigma \in A} \delta_{\lambda \in B} \left<\mu \nu | \frac{\partial \lambda}{\partial R_{B\beta}} \frac{\partial \sigma}{\partial R_{A\alpha}} \right>
		\end{aligned}
\end{equation}
where the notation has been simplified by just using the AO index.

\begin{equation}
		\begin{aligned}
			S^{(2)}_{\text{AO}~A\alpha B\beta, \mu\nu} =& \frac{\partial^2}{\partial R_{A\alpha} \partial R_{B\beta}} \braket{\chi_\mu}{\chi_\nu} \\ 
			=& \delta_{\mu \in A} \delta_{\mu \in B} \braket{\frac{\partial^2 \chi_\mu}{\partial R_{A\alpha} \partial R_{B\beta}}}{\chi_\nu} + \delta_{\nu \in A} \delta_{\nu \in B} \braket{\chi_\mu}{\frac{\partial^2 \chi_\nu}{\partial R_{A\alpha} \partial R_{B\beta}}} \\
			&+ \delta_{\mu \in A} \delta_{\nu \in B} \braket{\frac{\partial \chi_\mu}{\partial R_{A\alpha}}}{\frac{\partial \chi_\nu}{\partial R_{B\beta}}} + \delta_{\mu \in B} \delta_{\nu \in A} \braket{\frac{\partial \chi_\mu}{\partial R_{B\beta}}}{\frac{\partial \chi_\nu}{\partial R_{A\alpha}}} \\
		\end{aligned}
\end{equation}
where by integration by parts:
\begin{equation}
		\braket{\dfrac{\partial\chi_\mu}{\partial R_{A\alpha}}}
		{\dfrac{\partial\chi_\nu}{\partial R_{B\beta}}}
		= -\braket{\dfrac{\partial^2\chi_\mu}
			{\partial R_{A\alpha}\,\partial R_{A\beta}}}{\chi_\nu},
		\qquad \mu\in A,\;\nu\in B.
\end{equation}

In the OMO frame, the first-order derivative of the one-electron integral is
\begin{equation}
	\begin{aligned}h^{(1)}_{\text{OMO}~A\alpha} 
		&=h^{(1)}_{\mathrm{bare}~A\alpha} 
		+ T^{(1)}_{A\alpha} h_\text{MO} 
		+h_{\mathrm{MO}}\,(T^{(1)}_{A\alpha})^T \\
		&= C_{\mathrm{ref}}^T 
		\frac{\partial h_{\mathrm{AO}}}{\partial R_{A\alpha}} 
		C_{\mathrm{ref}}
		+ T^{(1)}_{A\alpha} h_\text{MO} 
		+h_{\mathrm{MO}}\,(T^{(1)}_{A\alpha})^T 
	\end{aligned}
	\label{eq:h-OMO}
\end{equation}
and for the two-electron integral
\begin{equation}
	\begin{aligned}
		g^{(1)}_{\text{OMO}~A\alpha, pqrs} 
		&= g^{(1)}_{\mathrm{bare}~ A\alpha, pqrs} \\
		&\quad + \sum_a T^{(1)}_{A\alpha, pa}\,g_{\mathrm{MO},aqrs}
		+ \sum_a T^{(1)}_{A\alpha, qa}\,g_{\mathrm{MO},pars} \\
		&\quad + \sum_a T^{(1)}_{A\alpha, ra}\,g_{\mathrm{MO},pqas}
		+ \sum_a T^{(1)}_{A\alpha, sa}\,g_{\mathrm{MO},pqra} \\
		&= \sum_{\mu\nu\kappa\lambda} 
		C_{\mu p}\,C_{\nu q}\,C_{\kappa r}\,C_{\lambda s} 
		\frac{\partial g_{\mathrm{AO},\mu\nu\kappa\lambda}}{\partial R_{A\alpha}} \\
		&\quad + \sum_a T^{(1)}_{A\alpha, pa}\,g_{\mathrm{MO},aqrs}
		+ \sum_a T^{(1)}_{A\alpha, qa}\,g_{\mathrm{MO},pars} \\
		&\quad + \sum_a T^{(1)}_{A\alpha, ra}\,g_{\mathrm{MO},pqas}
		+ \sum_a T^{(1)}_{A\alpha, sa}\,g_{\mathrm{MO},pqra} \\
		\label{eq:g-OMO}
	\end{aligned}
\end{equation}

\subsection{Orbital Hessian and Linear Response Demonstration}
By definition of the orbital Hessian:
\begin{equation}
	\frac{\partial^2 E}{\partial \theta_k\, \partial \theta_l}\bigg|_0
	= \langle 0 | \bigl[\hat{\Theta}_k,\, [\hat{\Theta}_l,\, \hat{H}]\bigr] | 0 \rangle
	= \frac{1}{2} \bra{0}  \bigl[\hat{\Theta}_k,\, [\hat{\Theta}_l,\, \hat{H}]\bigr]  \ket{0} +  \frac{1}{2} \bra{0}  \bigl[\hat{\Theta}_l,\, [\hat{\Theta}_k,\, \hat{H}]\bigr]  \ket{0}
\end{equation}
where we have used in the last equality the symmetry of the Hessian tensor. Moreover,
since $[\hat{\Theta}_l, \hat{H}] = -[\hat{H}, \hat{\Theta}_l]$, this is
equivalently:
\begin{equation}
	\frac{\partial^2 E}{\partial \theta_k\, \partial \theta_l}\bigg|_0
	= -\langle 0 | \bigl[\hat{\Theta}_k,\, [\hat{H},\, \hat{\Theta}_l]\bigr] | 0 \rangle.
	\label{eq:hessian-double-commutator}
\end{equation}

\subsection{Remaining Linear Response definitions}
The main LR matrices:
\begin{align}
 \textbf{E}^{[2]} &= \begin{pmatrix}
    {\boldsymbol{A}} & {\boldsymbol{B}} \\
      {\boldsymbol{B}}^* & {\boldsymbol{A}}^*           
     \end{pmatrix}, \quad 
     \textbf{S}^{[2]} = \begin{pmatrix}
    \boldsymbol{\Sigma} & \boldsymbol{\Delta} \\
     -\boldsymbol{\Delta} ^* &  -\boldsymbol{\Sigma}^*           
     \end{pmatrix}
\end{align}
The $\textbf{S}^{[2]}$ involves commutators between the orbital rotation and the active space excitation operators. 
\begin{align}
    \boldsymbol{\Sigma} & = \begin{pmatrix}
\left<0\left|\left[\hat{q}_\mu^\dagger,\hat{q}_{\nu}\right]\right|0\right>
& \left<0\left|\left[\hat{q}_{\mu}^\dagger,\hat{G}_{m}\right]\right|0\right> \\
\left<0\left|\left[\hat{G}_{n}^\dagger,\hat{q}_{\nu}\right]\right|0\right>
& \left<0\left|\left[\hat{G}_{n}^\dagger,\hat{G}_{m}\right]\right|0\right>
\end{pmatrix} \\
    \boldsymbol{\Delta} & = \begin{pmatrix}
\left<0\left|\left[\hat{q}_\mu^\dagger,\hat{q}_{\nu}^\dagger\right]\right|0\right>
& \left<0\left|\left[\hat{q}_{\mu}^\dagger,\hat{G}_{m}^\dagger\right]\right|0\right> \\
\left<0\left|\left[\hat{G}_{n}^\dagger,\hat{q}_{\nu}^\dagger\right]\right|0\right>
& \left<0\left|\left[\hat{G}_{n}^\dagger,\hat{G}_{m}^\dagger\right]\right|0\right>
\end{pmatrix}
\end{align}
The excitation vector is:
\begin{align}
    {\boldsymbol{\beta}}_k = \begin{pmatrix}
    {\boldsymbol{Z}}_k \\
      {\boldsymbol{Y}}_k^*
     \end{pmatrix} 
\end{align}
where the excitation block $\boldsymbol{Z}_k$ corresponds to positif frequencies $\omega_k >0$, and the de-excitation block $\boldsymbol{Y}_k^*$ corresponds to negative frequencies $\omega_k<0$.


\subsection{Methods and Ansatz}
\begin{figure}[H]
	\centering
	\includegraphics[width=1.0\columnwidth]{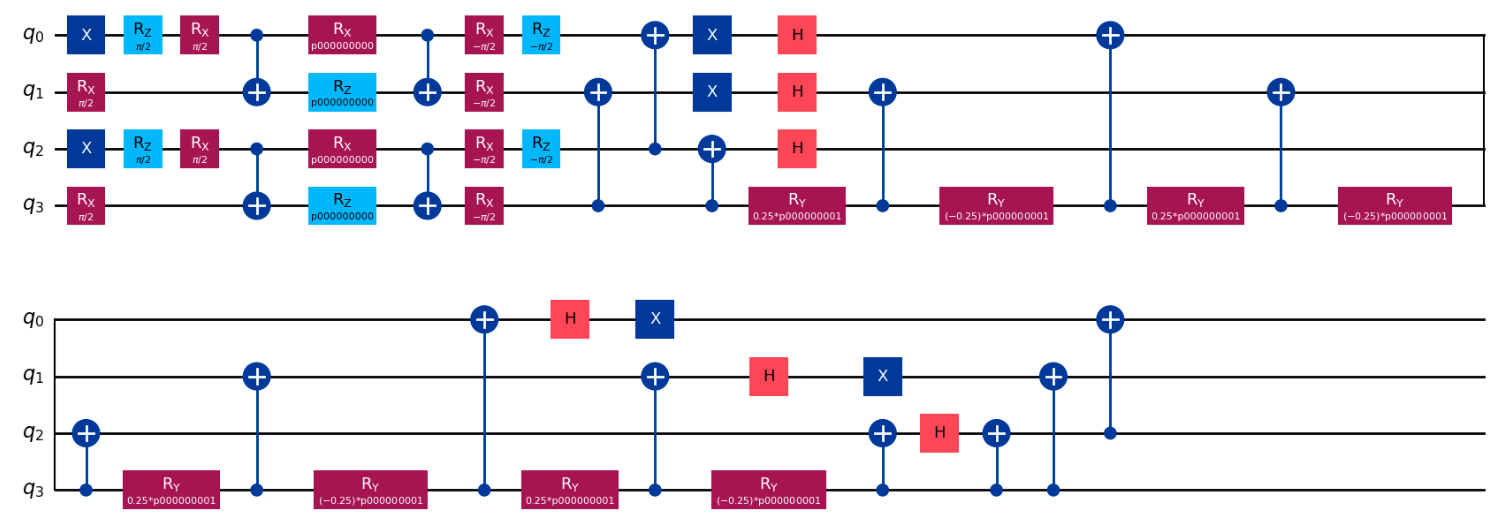}
	\caption{One layer pp-t-UPS/AS(2,2)/STO-3G circuit for the hydrogen molecule. }
	\label{fig:pptUPS_circuit}
\end{figure}

\begin{figure}[H]
	\centering
	\includegraphics[width=1.0\columnwidth]{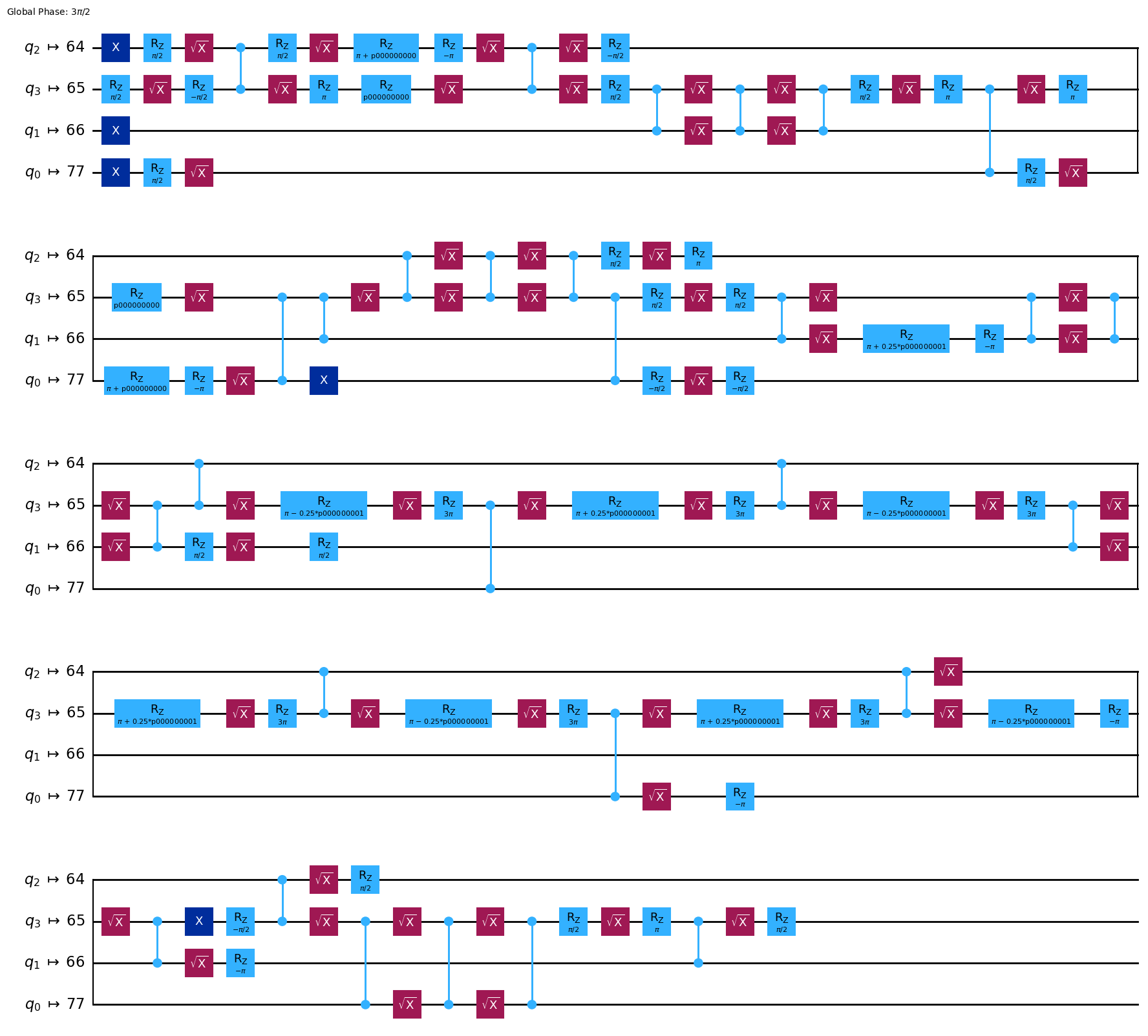}
	\caption{Transpiled one layer pp-t-UPS/AS(2,2)/STO-3G circuit for the hydrogen molecule. }
	\label{fig:pptUPS_circuit_transpiled}
\end{figure}

\begin{figure}[H]
	\centering
	\begin{tabular}{cccc}
		\hline
		\textbf{Qubit} & \textbf{Readout Assignment Error} & \textbf{Measure Error} & \textbf{T1 ($\mu$s)} \\
		\hline
		54 & 0.1179 & 0.1179 & 336.3 \\
		\hline
		23 & 0.0962 & 0.0962 & 240.5 \\
		\hline
		9 & 0.0818 & 0.0818 & 95.2 \\
		\hline
		117 & 0.0803 & 0.0803 & 7.2 \\
		\hline
		72 & 0.0576 & 0.0576 & 307.7 \\
		\hline
		140 & 0.0566 & 0.0566 & 177.3 \\
		\hline
		21 & 0.0520 & 0.0520 & 426.4 \\
		\hline
		36 & 0.0378 & 0.0378 & 5.8 \\
		\hline
		32 & 0.0330 & 0.0330 & 325.7 \\
		\hline
		83 & 0.0325 & 0.0325 & 105.6 \\
		\hline
	\end{tabular}
	\includegraphics[width=1.0\columnwidth]{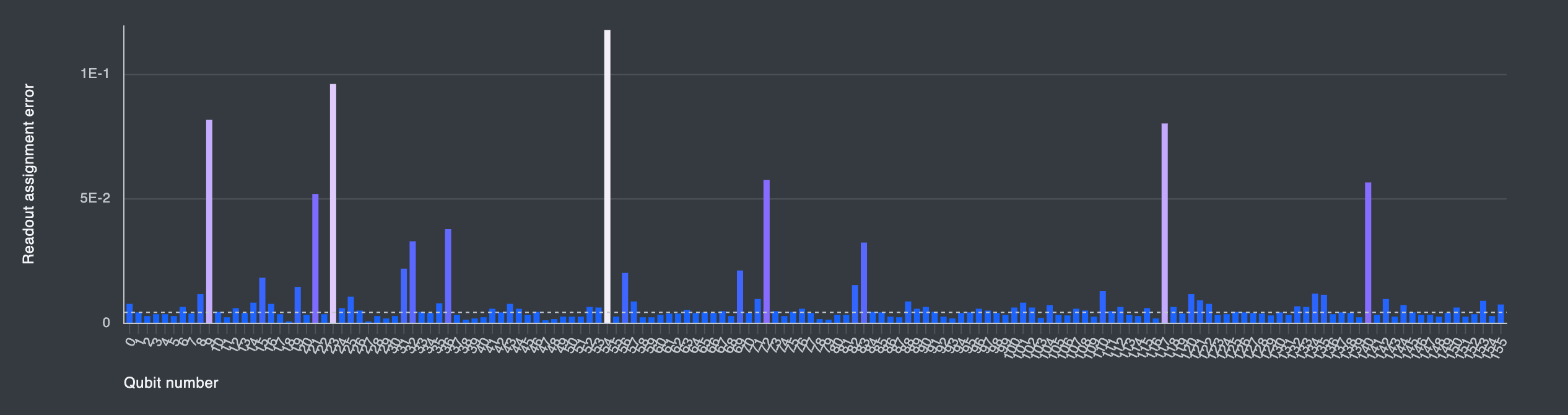}
	\caption{(Top) The 10 most unreliable qubits on IBM's Pittsburgh quantum computer. The $T_1$ value correspond to the coherence time. (Bottom)  Read out error calibration graph of every qubit. Any measurement using any  of this qubits is dropped}
	\label{fig:qubit_error}
\end{figure}

\subsection{Gradient Results}

\clearpage
\subsection{Hessian Results}

\newcolumntype{C}[1]{>{\centering\arraybackslash}p{#1}}

\newcommand{\mstd}[2]{%
  \begin{tabular}[t]{@{}c@{}}
    $#1$ \\[-2pt]
    {\footnotesize$\pm #2$}
  \end{tabular}}

\newcommand{\mmean}[1]{$#1$}

\begin{table}[htbp]
  \centering
  \caption{1-RDM (Reduced Density Matrix) element-wise statistics
           (mean $\pm$ std, 46 samples).}
  \label{tab:1rdm}
  \begin{minipage}[t]{0.45\linewidth}
    \centering
    \subcaption{Ideal tUPS reference}
    \begin{tabular}{C{2.6cm}C{2.6cm}}
      \toprule 
      \mmean{1.9677}                   & \mmean{0.11974} \\[14pt]
      \mmean{0.11974}      & \mmean{0.03225} \\[14pt]
      \bottomrule
    \end{tabular}
  \end{minipage}%
  \hfill
  \begin{minipage}[t]{0.45\linewidth}
    \centering
    \subcaption{QPU tUPS experiment}
    \begin{tabular}{C{2.6cm}C{2.6cm}}
      \toprule
      \mstd{1.96060}{0.00634}              & \mstd{0.10578}{0.00656} \\[14pt]
      \mstd{0.10578}{0.00656} & \mstd{0.03941}{0.00634} \\[14pt]
      \bottomrule
    \end{tabular}
  \end{minipage}
\end{table}

\begin{table}[htbp]
  \centering
  \caption{Linear Response A matrix element-wise statistics
           (mean $\pm$ std, 46 samples).}
  \label{tab:lra}
  \begin{minipage}[t]{0.45\linewidth}
    \centering
    \subcaption{Ideal tUPS reference}
    \begin{tabular}{C{2.6cm}C{2.6cm}}
      \toprule
      \mmean{0.97016} & \mmean{0.11231} \\[14pt]
      \mmean{0.11231} & \mmean{1.5758}              \\[14pt]
      \bottomrule
    \end{tabular}
  \end{minipage}%
  \hfill
  \begin{minipage}[t]{0.45\linewidth}
    \centering
    \subcaption{QPU tUPS experiment}
    \begin{tabular}{C{2.6cm}C{2.6cm}}
      \toprule
      \mstd{0.95881}{0.00457} & \mstd{0.09857}{0.00593} \\[14pt]
      \mstd{0.09857}{0.00593} & \mstd{1.56060}{0.00920}              \\[14pt]
      \bottomrule
    \end{tabular}
  \end{minipage}
\end{table}

\begin{table}[htbp]
	\centering
	\caption{Linear Response B matrix element-wise statistics
		(mean $\pm$ std, 46 samples).}
	\label{tab:lrb}
	\begin{minipage}[t]{0.45\linewidth}
		\centering
		\subcaption{Ideal tUPS reference}
		\begingroup\large
		\begin{tabular}{C{3.0cm}C{3.0cm}}
			\toprule
			\mmean{-0.22054} & \mmean{0.02978}  \\[14pt]
			\mmean{0.02978}  & \mmean{-0.03789} \\[14pt]
			\bottomrule
		\end{tabular}
		\endgroup
	\end{minipage}%
	\hfill
	\begin{minipage}[t]{0.45\linewidth}
		\centering
		\subcaption{QPU tUPS experiment}
		\begingroup\large
		\begin{tabular}{C{3.0cm}C{3.0cm}}
			\toprule
			\mstd{-0.21601}{0.00169} & \mstd{0.02699}{0.00105}  \\[14pt]
			\mstd{0.02699}{0.00105}  & \mstd{-0.03505}{0.00160} \\[14pt]
			\bottomrule
		\end{tabular}
		\endgroup
	\end{minipage}
\end{table}

\begin{table}[htbp]
  \centering
  \caption{2-RDM element-wise statistics
           (mean $\pm$ std, 46 samples).}
  \label{tab:2rdm}

  \subcaption{Ideal tUPS reference}
  \resizebox{\textwidth}{!}{%
  \begin{tabular}{C{2.2cm}C{2.2cm}C{2.2cm}C{2.2cm}}
    \toprule
    \mmean{1.9585}                  & \mmean{0.13431}  & \mmean{0.13431}  & \mmean{0.00921}  \\[6pt]
    \mmean{0.13431}     & \mmean{-0.21243} & \mmean{0.00921}  & \mmean{-0.01457} \\[6pt]
    \mmean{0.13431}     & \mmean{0.00921}  & \mmean{-0.21243} & \mmean{-0.01457} \\[6pt]
    \mmean{0.00921}     & \mmean{-0.01457} & \mmean{-0.01457} & \mmean{0.02304}  \\[6pt]
    \bottomrule
  \end{tabular}}

  \bigskip
  \subcaption{QPU tUPS experiment}
  \resizebox{\textwidth}{!}{%
  \begin{tabular}{C{2.2cm}C{2.2cm}C{2.2cm}C{2.2cm}}
    \toprule
    \mstd{1.95050}{0.00709}               & \mstd{0.11759}{0.00706}  & \mstd{0.11759}{0.00706}  & \mstd{0.01007}{0.00127}  \\[14pt]
    \mstd{0.11759}{0.00706}  & \mstd{-0.19650}{0.00899} & \mstd{0.00913}{0.00422}  & \mstd{-0.01268}{0.00233} \\[14pt]
    \mstd{0.11759}{0.00706}  & \mstd{0.00913}{0.00422}  & \mstd{-0.19650}{0.00899} & \mstd{-0.01268}{0.00233} \\[14pt]
    \mstd{0.01007}{0.00127}  & \mstd{-0.01268}{0.00233} & \mstd{-0.01268}{0.00233} & \mstd{0.02934}{0.00576}  \\[14pt]
    \bottomrule
  \end{tabular}}
\end{table}

\begin{table}[htbp]
  \centering
  \caption{Nuclear Hessian matrix element-wise statistics
           (mean $\pm$ std, 46 samples). The elements are in Ha/Bohr$^{-2}$}
  \label{tab:hessian}

  \subcaption{FCI Finite Difference}
  \resizebox{\textwidth}{!}{%
  \begin{tabular}{C{2.2cm}C{2.2cm}C{2.2cm}C{2.2cm}C{2.2cm}C{2.2cm}}
    \toprule
    \mmean{0.00002}  & \mmean{0} & \mmean{0} & \mmean{-0.00002} & \mmean{0} & \mmean{0} \\[14pt]
    \mmean{0} & \mmean{0.00002}  & \mmean{0} & \mmean{0} & \mmean{-0.00002} & \mmean{0} \\[14pt]
    \mmean{0} & \mmean{0} & \mmean{0.47702}  & \mmean{0} & \mmean{0} & \mmean{-0.47702}  \\[14pt]
    \mmean{-0.00002} & \mmean{0} & \mmean{0} & \mmean{0.00002}  & \mmean{0} & \mmean{0} \\[14pt]
    \mmean{0} & \mmean{-0.00002} & \mmean{0} & \mmean{0} & \mmean{0.00002}  & \mmean{0} \\[14pt]
    \mmean{0} & \mmean{0} & \mmean{-0.47702}  & \mmean{0} & \mmean{0} & \mmean{0.47702}   \\[14pt]
    \bottomrule
  \end{tabular}}

  \bigskip
  \subcaption{Ideal tUPS reference}
  \resizebox{\textwidth}{!}{%
  \begin{tabular}{C{2.2cm}C{2.2cm}C{2.2cm}C{2.2cm}C{2.2cm}C{2.2cm}}
    \toprule
    \mmean{0.00002}  & \mmean{0} & \mmean{0} & \mmean{-0.00002} & \mmean{0} & \mmean{0} \\[14pt]
    \mmean{0} & \mmean{0.00002}  & \mmean{0} & \mmean{0} & \mmean{-0.00002} & \mmean{0} \\[14pt]
    \mmean{0} & \mmean{0} & \mmean{0.47554}  & \mmean{0} & \mmean{0} & \mmean{-0.47554} \\[14pt]
    \mmean{-0.00002} & \mmean{0} & \mmean{0} & \mmean{0.00002}  & \mmean{0} & \mmean{0} \\[14pt]
    \mmean{0} & \mmean{-0.00002} & \mmean{0} & \mmean{0} & \mmean{0.00002}  & \mmean{0} \\[14pt]
    \mmean{0} & \mmean{0} & \mmean{-0.47554} & \mmean{0} & \mmean{0} & \mmean{0.47554}  \\[14pt]
    \bottomrule
  \end{tabular}}

  \bigskip
  \subcaption{QPU tUPS experiment}
  \resizebox{\textwidth}{!}{%
  \begin{tabular}{C{2.0cm}C{2.0cm}C{2.0cm}C{2.0cm}C{2.0cm}C{2.0cm}}
    \toprule
    \mstd{-0.00371}{0.00301} & \mstd{0}{0} & \mstd{0}{0} & \mstd{0.00371}{0.00301}  & \mstd{0}{0} & \mstd{0}{0} \\[14pt]
    \mstd{0}{0} & \mstd{-0.00371}{0.00301} & \mstd{0}{0} & \mstd{0}{0} & \mstd{0.00371}{0.00301}  & \mstd{0}{0} \\[14pt]
    \mstd{0}{0} & \mstd{0}{0} & \mstd{0.47841}{0.00405}  & \mstd{0}{0} & \mstd{0}{0} & \mstd{-0.47841}{0.00405} \\[14pt]
    \mstd{0.00371}{0.00301}  & \mstd{0}{0} & \mstd{0}{0} & \mstd{-0.00371}{0.00301} & \mstd{0}{0} & \mstd{0}{0} \\[14pt]
    \mstd{0}{0} & \mstd{0.00371}{0.00301}  & \mstd{0}{0} & \mstd{0}{0} & \mstd{-0.00371}{0.00301} & \mstd{0}{0} \\[14pt]
    \mstd{0}{0} & \mstd{0}{0} & \mstd{-0.47841}{0.00405} & \mstd{0}{0} & \mstd{0}{0} & \mstd{0.47841}{0.00405}  \\[14pt]
    \bottomrule
  \end{tabular}}
\end{table}


    


\clearpage

\printbibliography


\end{document}